# Simulation of Crowd Egress with Environmental Stressors

**Peng Wang** **Xiaoda Wang** **Peter Luh**

**Christian Wilkie** **Timo Korhonen** **W. M. Wonham** **Neal Olderman**

This article introduces a modeling framework to characterize evacuee response to environmental stimuli during emergency egress. The model is developed in consistency with stress theory, which explains how an organism reacts to environmental stressors. We integrate the theory into the well-known social-force model, and develop a framework to simulate crowd evacuation behavior in multi-compartment buildings. Our method serves as a theoretical basis to study crowd movement at bottlenecks, and further simulate their herding and way-finding behavior in normal and hazardous conditions. The pre-movement behavior is also briefly investigated by using opinion dynamics. The algorithms have been partly tested in FDS+EVAC as well as our simulation platform crowdEgress .

***Glossary—***

**Crowd Egress:** A multitude of people with conscious mind, and they move and interact collectively towards a place of safety. In this article the term mainly refers to an evacuation process in a multi-compartment layout of buildings.

**Stress**: The word stress relates to a mental strain or a harmful environmental stimuli that cause response of organism. The current usage of this term was mainly referred to the interaction between the environment and the living bodies, emphasizing the role of the individual's appraisal of situations in shaping their responses.

**Social Force Model:** Force-based model of social interaction of individuals, and it is based on the Langiven equation, and it enables agent-based simulation of emergent behavior of a large number of people in normal and emergency situation.

**Herding Effect:** When people are responding to an emergency or disaster event, they are more inclined to follow each other and behave as a social group rather than independently in respond to the outside. This is a base instinct, not only for human crowd, but also for many social animals, e.g., bird flock, sheep herd and fish school.

**Vicsek Model:** A physics-based version of Boids flocking model, where every boid aligns its moving direction along the average direction of all the boids in its interaction range. Noise is commonly added in the direction of the boid in the sense of statistical physics.

**Opinion and Decision Process:** Mathematical modeling of opinion exchange or decision making process in which individuals interact and influence each other in a specific social context.

**Way-Finding Activities:** Route choice of evacuees in which individuals decide on their walking path between two locations based on the environmental and social context, and it is a fundamental problem cross disciplines.

---

Peng Wang previously studied in the Department of Electrical and Computer Engineering, University of Connecticut, Storrs. The research work was supported by NSF Grant # CMMI- 1000495.

# 1. Social Force Model and Stress Theory

Historically, stress was initially a physical quantity, which is a measure of the internal forces in a body between its particles. In the 1920s and 1930s, biological and psychological circles occasionally used the term to refer to a mental strain or a harmful environmental stimuli that cause response of organism. The current usage of the word stress was mainly referred to the interaction between environmental stimuli and the living bodies, emphasizing the role of the individual's appraisal of situations in shaping their responses (Selye, 1975).

For human beings stress is normally perceived when we think the demand being placed on us exceed our ability to cope with, and it can be external and related to the environment, and it becomes effective by internal perception. This paper will integrate the stress theory into the well-known social force model (Helbing et. al., 2002 and 2005), which has been widely used in simulation of crowd evacuation in the past two decades. The feasibility of our method relies on certain psychological factors abstracted in the model, such as "desired velocity" and "social force." For instance, desired velocity is not the commonly-known velocity in physics, but refers to the moving speed and direction that an individual expects to realize, and it actually exists in one's mind, and it could be sometimes equal to physical velocity, but conceptually they are different. This new concept enables us to model how the motivation level in the psychological sense (i.e., desired velocity) leads to behavioral change in the physics world (i.e., actually velocity), and such motivation level are result of human perception, and are adapted to the environmental stressors. Thus, it is feasible to extend social-force model to characterize individuals' response to their surroundings based on stress theory. As below we present a diagram to describe the interplay between individuals and their surroundings based on the extended social-force model.

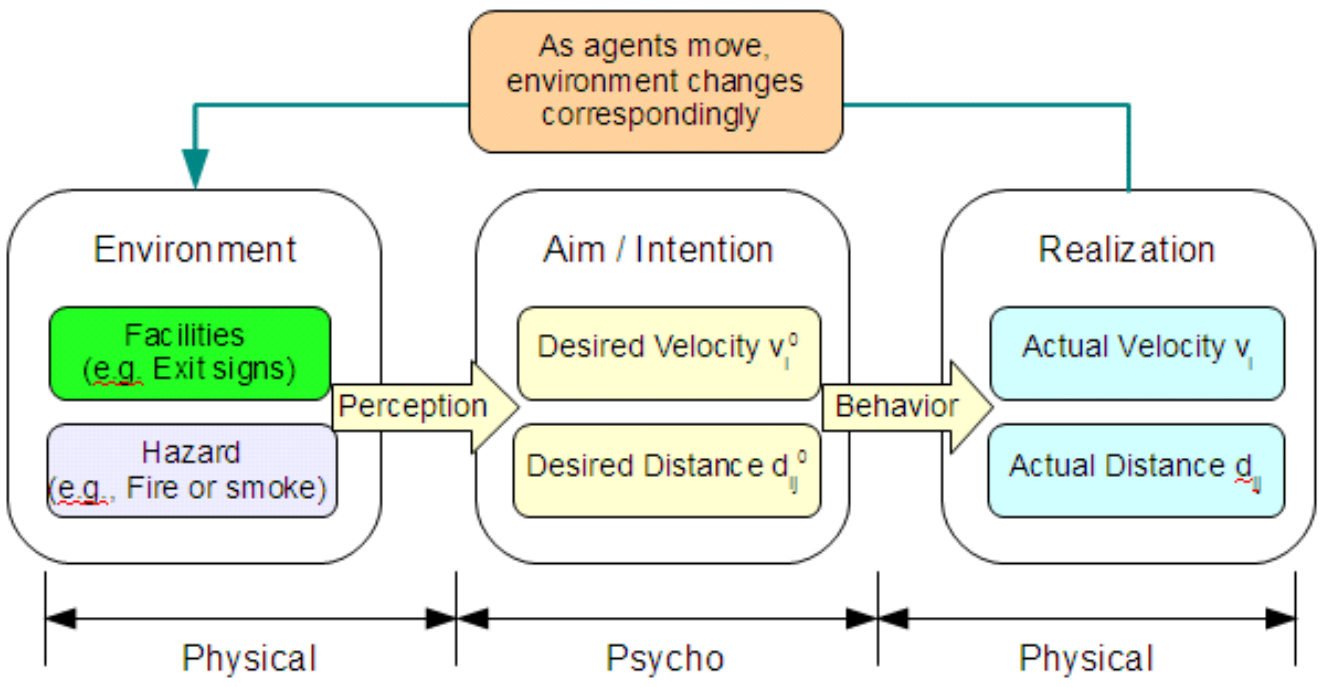

**Figure 1.1. Perception and Behavior in a Feedback Loop: The motivation level $v_i^0$ and $d_{ij}^0$ in social force model are the result of human perception, and are adapted to the environmental stressors such as fire and smoke, and $v_i^0$ and $d_{ij}^0$ could vary both temporally and spatially, and they lead to behavior change in $v_i$ and $d_{ij}$. The social-force model is extended to characterize the interplay between individuals and their surroundings.**

In the above diagram the stressors include egress facilities (e.g., alarm, guidance) and hazard (e.g., fire and smoke), and people respond to such environmental stressors, moving to the safety. In a psychological sense such stress is caused by mismatch between psychological demand and realistic situation (Staal, 2004), and we summarize the mismatch in terms of velocity and interpersonal distance as listed in Table 1: the psychological demands are abstracted as desired velocity $\boldsymbol{v}_i^0$ and desired interpersonal distance $d_{ij}^0$ while the physical reality is described by the physical velocity $\boldsymbol{v}_i$ and distance $d_{ij}$. The difference of two variables measures how much people feel stressed, by which people are motivated into certain behavior. The motivation is abstracted as the driving force and social force as shown in Table 1. The entire process formalizes the stimuli-reaction model (S-R model) in psychological view of behaviorism.

Table 1.1 On Conception of Stress in Extended Social-Force Model

| | *Opinion (Psychological Characteristics)* | *Behavior (Physic-Based Characteristics)* | *Difference between subjective opinion and objective reality* | *Forced-Based Term for Newton Second Law* |
|---|---|---|---|---|
| ***Time-Related Stress: Velocity*** | desired velocities $\boldsymbol{v}_i^0$ | actual velocities $\boldsymbol{v}_i$ | ***Time-Related Stress: Velocity*** $\boldsymbol{v}_i^0 - \boldsymbol{v}_i$ or $\lvert\boldsymbol{v}_i\rvert / \lvert\boldsymbol{v}_i^0\rvert$ | ***Driving Force*** $\boldsymbol{f}^{drv} = m_i(\boldsymbol{v}_i^0 - \boldsymbol{v}_i)/\tau$ |
| ***Space-related Stress: Distance*** | desired distance $d_{ij}^0$ | actual distance $d_{ij}$ | ***Space-related Stress: Distance*** $d_{ij}^0 - d_{ij}$ or $d_{ij}/d_{ij}^0$ | ***Social Force*** $\boldsymbol{f}_{ij}^{soc} = A_i exp((d_{ij}^0 - d_{ij})/B_i)\boldsymbol{n}_{ij}$ |

There are two types of stress listed in Table 1. The first type refers to time-related stress which is commonly known as time-pressure, and it is measured by the difference of desired velocity and actual velocity, i.e., $\boldsymbol{v}_i^0 - \boldsymbol{v}_i$. The second type indicates space-related stress, which refers to proxemics and social norms, and is represented by the gap of desired interpersonal distance and actual interpersonal distance, i.e, $d_{ij}^0 - d_{ij}$.

Usually emergency egress is a stressful situation. When people hear the fire alarm and confirm any kind of threat in their surroundings, they normally begin to escape, and this process is called movement phase in crowd egress. Each individual evacuee's motion is governed by physics laws, and the motion equation of individual $i$ is specified as below.

$$m_i \frac{d\boldsymbol{v}_i(t)}{dt} = \boldsymbol{f}_i^{drv} + \sum_{j(\neq i)} \boldsymbol{f}_{ij} + \sum_{w} \boldsymbol{f}_{iw} + \sum_{h} \boldsymbol{f}_{ih} \quad , \tag{1.1}$$

The mass of an individual is denoted by $m_i$. The change of the instantaneous velocity $\boldsymbol{v}_i(t)$ of individual $i$ is determined by the total force. The forces motivates people to adapt to surrounding stimuli, and they consist of driving force $\boldsymbol{f}_i^{drv}$, interaction force $\boldsymbol{f}_{ij}$ as well as boundary wall force $\boldsymbol{f}_{iw}$. The hazard impact is also abstracted as a kind of force $\boldsymbol{f}_{ih}$.

The driving force functions as an energy source that drives people to move to a destination exit, and it is commonly specified by $\boldsymbol{f}^{drv} = \boldsymbol{F}^{drv}(\boldsymbol{v}_i^0 - \boldsymbol{v}_i)$, where $\boldsymbol{F}^{drv}(.)$ is generally a monotonically increasing function. The interaction force $\boldsymbol{f}_{ij}$ describes the social-psychological tendency of two individuals to keep proper distance (as called the social-force), and if people have physical contact with each other, physical forces are also taken into account, i.e., $\boldsymbol{f}_{ij} = \boldsymbol{f}_{ij}^{soc} + \boldsymbol{f}_{ij}^{phy}$. In this article we mainly focus on the social force term. In Helbing et. al., 2002 and 2005 this term is given in an exponential form, and we will integrate a new concept of desired distance $d_{ij}^0$ into this force term. In a general sense we specify this force term by $\boldsymbol{f}_{ij}^{soc} = \boldsymbol{F}^{soc}(d_{ij}^0 - d_{ij})$. The interaction of an individual with obstacles like walls is treated analogously, and denoted by $\boldsymbol{f}_{iw} = \boldsymbol{f}_{iw}^{soc} + \boldsymbol{f}_{iw}^{phy}$, and this force specifies boundary of people's motion. The hazard force is a special term we take into account for emergency egress and we will briefly discuss it in detail soon later.

Next we will emphasize that $\boldsymbol{f}_i^{drv}$ $\boldsymbol{f}_{ij}^{soc}$ $\boldsymbol{f}_{iw}^{soc}$ and $\boldsymbol{f}_{ih}$ are all subjective entities coming from people's opinions, and they are generated intentionally by people through foot-ground friction on physics basis. These forces essentially describe how an individual perceive and react to the outside environmental stimuli. Because $\boldsymbol{f}_{iw}^{soc}$ can be formulated in analog of $\boldsymbol{f}_{ij}^{soc}$, we will not further discuss $\boldsymbol{f}_{iw}^{soc}$ in this article. In sum, $\boldsymbol{v}_i^0$ and $d_{ij}^0$ are non-physics entities for they exist in people's opinion, not in the physical world, and we have well explained the psychological background of the model in Wang, 2016, and critically modified the key concept of social force in consistency of both physical laws and psychological principles.

In the following section we will mainly introduce $\boldsymbol{f}_i^{drv}$ $\boldsymbol{f}_{ij}^{soc}$ and $\boldsymbol{f}_{ih}$ in the background of evacuation study. Our focus is applying the above model in simulation of crowd egress, especially in modeling how perception of hazard (fire and smoke) and guidance (e.g., exit signs) influence evacuees' escape behavior. The method has been partly tested in FDS+Evac, a well-known open-source evacuation simulator (Korhonen, 2017; Forney 2017; McGrattan, 2018; Hostikka et. al., 2007; Korhonen et. al., 2008) as well as our simulation platform CrowdEgress (Wang et. al., 2023).

Last but not least, an evacuation process is commonly divided into two phases: pre-movement phase and movement phase. In the pre-movement phase evacuees are inclined to collect information rather than actually move to any exit. The pre-movement behavior will be also discussed in this article, where existing model in opinion dynamics are applied to calculate pre-movement time of many individuals. We will especially introduce this model in Section 2 and 3.

### (a) Adapting Desired Velocity To Environmental Stressors: Fight-or-Flight Response and Hazard Effect

When the fire/smoke spread towards people, people normally desire moving faster to escape from danger (Proulx , 1993; Ozel, 2001; Kuligowski, 2009). Thus, we suggest that the desired velocity $\boldsymbol{v}^0$ increase when people perceive such danger, and this means increase of arousal, resulting in fight-or-flight response in psychological theory (Cannon, 1932). As a result the driving force is increased, and people are motivated to respond and move to the safety. This process suggests that people transfer their internal energy into kinetic form in order to realize desired motion. Such driving force depends on how much an individual feel stressed and thus increase with $\boldsymbol{v}_i^0 - \boldsymbol{v}_i$.

A widely-used formula of the driving force is given in linear form $\boldsymbol{f}^{drv} = m_i(\boldsymbol{v}_i^0 - \boldsymbol{v}_i)/\tau$, and it describes an individual tries to move with a desired velocity $\boldsymbol{v}_i^0$ and expects to adapt the actual velocity $\boldsymbol{v}_i$ to the desired velocity $\boldsymbol{v}_i^0$ within a certain time interval. In specific $\boldsymbol{v}_i^0$ is the target velocity existing in one's opinion while $\boldsymbol{v}_i$ is the physical velocity being achieved in the reality. Thus, $\boldsymbol{v}_i^0 - \boldsymbol{v}_i$ implies the difference between the human subjective wish and realistic situation, and it is scaled by a time parameter $\tau_i$ to formulate an acceleration term, making the physical velocity $\boldsymbol{v}_i$ approaching towards desired velocity $\boldsymbol{v}_i^0$. The convergence rate is determined by $m_i/\tau$. If $\boldsymbol{v}_i^0$ changes with time, $\boldsymbol{v}_i$ will approach to $\boldsymbol{v}_i^0$ with a time delay. This mathematical description of driving force was initially used for vehicle traffic flow problem (Payne, 1971; Whitham, 1974), and it was also applied to pedestrian traffic problem in Helbing et. al., 2002 and 2005.

Interestingly, velocity is a time-related concept in physics and thus $\boldsymbol{v}_i^0 - \boldsymbol{v}_i$ reflects a kind of time-related stress. Such stress usually relates to a time-critical situation, or so-called "emergency." This may explains why the model is suitable for simulation of emergency evacuation in the past twenty years. Consequently the driving force is a function of $\boldsymbol{v}_i^0 - \boldsymbol{v}_i$, which motivate people to move to a destination exit. In this modeling method the desired velocity $\boldsymbol{v}_i^0$ indicates one's arousal level, which relates to one's awareness of the situation, and $\boldsymbol{v}_i^0$ increases when an individual directly detects the harmful gas or smoke, or indirectly informed by the fire alarms.

However, increasing $\boldsymbol{v}_i^0$ does not mean that individuals are able to move faster, especially in the movement towards an exit. There is possibility that people may slow down when passing through a bottleneck or a hazard-filled area (e.g., smoke-filled corridors). As mathematically described by Equation (1.1), the impact of hazard is indicated by a resistance force which increases with hazard intensity (e.g., smoke density). This hazard force describes how hazardous condition impedes people's motion in egress, and it is denoted by $\boldsymbol{f}_{ih}$. In the following discussion we usually take smoke for example and $\boldsymbol{f}_{ih}$ is supposed to be a function of smoke density. Other hazard characteristics can also be considered such as gas temperature. In a sense Equation (1.1) indicates that the hazard condition (e.g., smoke) is a kind of "spreading walls" that impede people's' motion. The force from solid walls are expressed by $\boldsymbol{f}_{iw}$ while force from hazardous condition are denoted by $\boldsymbol{f}_{ih}$. An individual is able to go through such "spreading walls" if the smoke is not thick. As smoke density increase, people are impeded and cannot get into that area (See Figure 1.2).

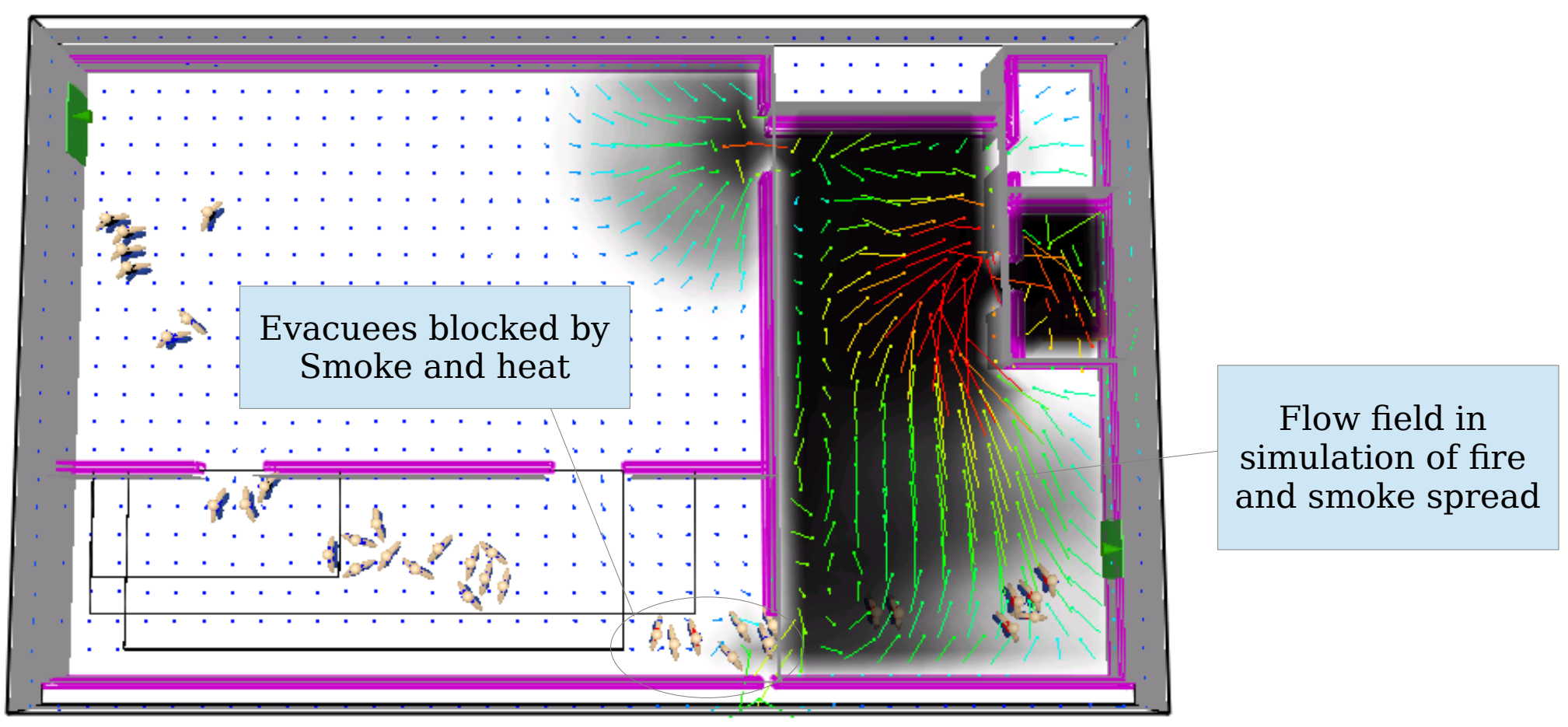

**Figure 1.2. Simulation of Crowd Evacuation with Smoke: Smoke spreads and it is like "moving walls" which block evacuees' movement, and evacuees are not able to pass through such "moving walls" if the smoke is thick. The testing results are obtained by modifying the source code of FDS+Evac (Version 6.5.3), and the data can be downloaded at https://github.com/crowdmodel/test-crowd-dynamics**

How to select the direction of $\boldsymbol{f}_{ih}$ is an interesting topic, and it mainly depends on characteristics of hazard. The physics law of smoke transport is not the same as that of heat transfer. A common method is assuming $\boldsymbol{f}_{ih}$ always impedes an evacuee's movement in any direction, and thus $\boldsymbol{f}_{ih}$ is always opposite to the direction of moving velocity $\boldsymbol{v}_i$. In FDS+Evac we use (-HR%U, -HR%V) to specify the direction of $\boldsymbol{f}_{ih}$, especially for resistance of smoke, and HR%U and HR%V compose 2D vector for human velocity. Another option is using gradient of hazard intensity. This gradient is useful to represent the direction of heat flux. The gradient points in the direction of the greatest increasing rate of hazard intensity. For example the hazard intensity is described by gas temperature $TMP_G(x, y)$ in a 2-dimensional plane at people's average height, and thus the direction of $\boldsymbol{f}_{ih}$ is opposite to the gradient of $TMP_G(x, y)$, which points in the direction of the greatest decreasing rate of hazard intensity.

$$\text{direction}(\boldsymbol{f}_{ih}) = -\,\text{direction}(\boldsymbol{v}_i) \qquad\qquad \text{direction}(\boldsymbol{f}_{ih}) = -\nabla TEM_G(\text{x, y}) \tag{1.2}$$

In sum the driving force and hazard force are conflicting factors, and they function together to give a whole picture of the model. The driving force motivates people to escape, describing how people are motivated into escape motion. In contrast the hazard force is used to describe if the outside condition permits such change or not. The following plot in Figure 1.3 exemplifies an increasing curve of the driving force component due to desired velocity $(m_i \boldsymbol{v}_i^0)/\tau$ and hazard resistance due to the smoke density. In Figure 1.3 it is noted that $\boldsymbol{v}_i^0$ increases in square root rate with smoke density (i.e., SOOT). and the hazard increases in quadratic rate. When the smoke density increases initially, people are able to speed up in smokiness (i.e., the left-side of the red line). As the smoke density keeps increasing, the resistance from smoke becomes more and more significant, and it is dominant to slow down evacuees' motion (i.e., the right-side of the red line). As real-world fire drills or experiments have suggested, such motion is mainly due to poor visibility on the path (Jin and Yamada, 1989; Fridolf et. al., 2013; Was, 2018) or reduced percentage of oxygen.

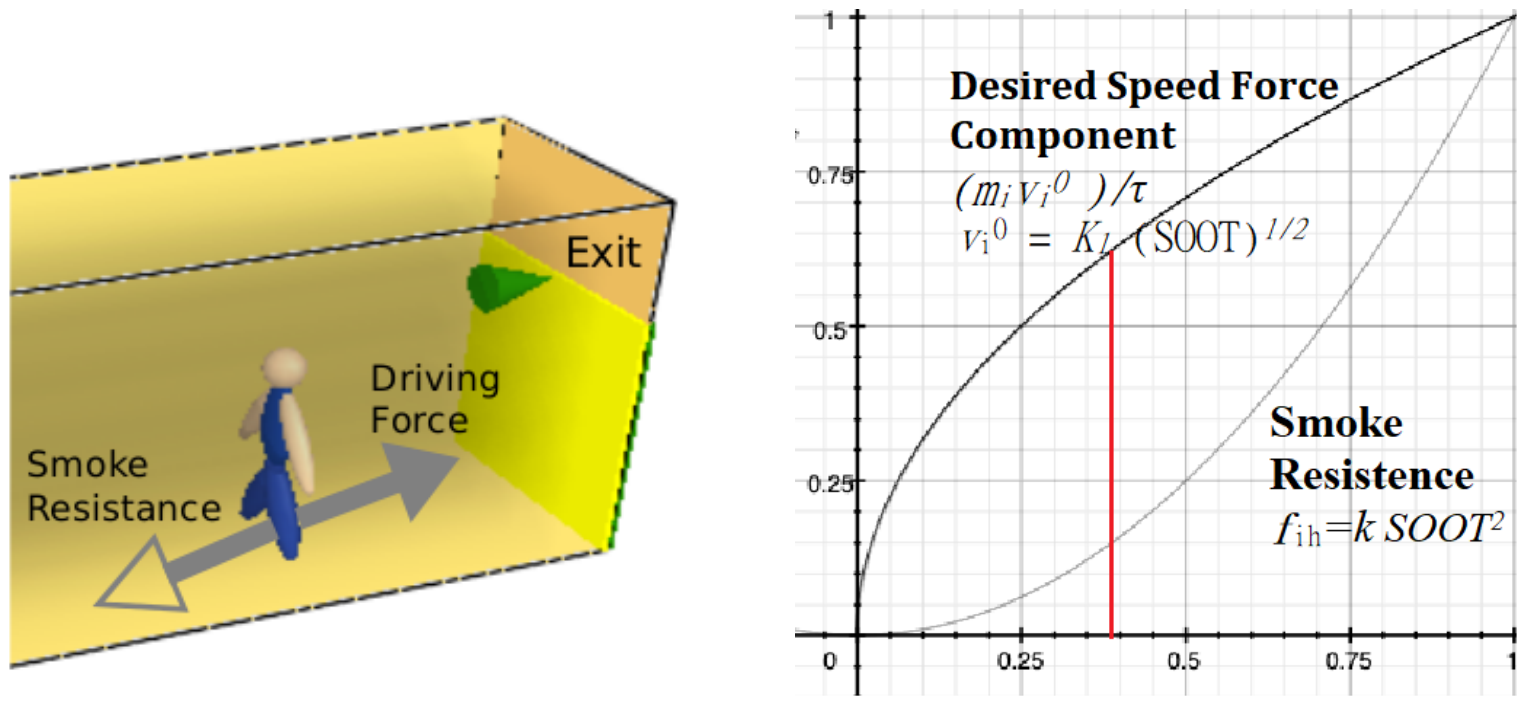

**Figure 1.3. Walking Behavior in Smoke Condition: When the smoke density increases initially, the smoke is not thick so that people are able to speed up. As the smoke density keeps increasing, the resistance from smoke is predominant and people have to slow down even if they desire moving fast in escape.**

The difficulty of the above method exists in quantitative analysis. It is not easy to quantify how fast evacuees desire moving or how they perceive threat from smoke or heat. Especially, the hazard force is not commonly a physics-based force, but a subjective entity which is closely related to human perception and cognition. In other words this force does not only depends on the hazard intensity itself, but also how people perceive it. For example, some people are sensitive to smoke inhaled while others may tolerate much. This effect may be inferred from certain clinical database, and theory in psychophysics may contribute valuable viewpoint (Stevens, 1971) because it discusses the relationship of the physics-based things and human perception of things. In brief, hazard force describes human perception of hazardous environmental stimuli. Although we call it "force" in the above model, it is actually different from the common "force" concept in physics, but related more about study subjects in psychology and psychophysics.

However, simulation is still a useful tool for risk analysis of building egress (Gwynne et. al., 2001, Sidiropoulos, Kiourt and Moussiades, 2020). It is relatively easy to adjust parameters and observe different scenarios. For instance, from the above simulation we learn how smoke affects people's escape behavior. There are standard examples of FDS+Evac to test walking speed of evacuees in smoke condition. Please refer to the section of supplementary data for details.

### (b) Adapting Desired Distance To Environmental Stressors: Proximics and Social Norms

In social psychology social norms are "representations of appropriate behavior" in a certain situation or environment, and it partly refers to a theory of how people use their personal space to interact with surrounding people. In Hall, 1963 the theory was named by proxemics, and it suggests that we surround ourselves with a "bubble" of personal space, which protects us from too much arousal and helps us keep comfortable when we interact with others. People normally feel stressed if such space is compressed. This theory justifies the repulsive social force in Helbing et. al., 2002 and 2005, where the force describes a potential field, and it reflects social repulsion of interacting individuals.

From the perspective of crowd modeling, the proxemics is represented by desired interpersonal distance $d_{ij}^0$ in our model, and $d_{ij}^0 - d_{ij}$ indicates a kind of space-related stress as shown in Table 1.1. Usually $d_{ij}^0$ changes with locations. For example, in elevators, entrance or narrow corridors, people usually accept smaller proximal distance, implying that desired interpersonal distance $d_{ij}^0$ is accepted to be small in these places. In sum adaption of $d_{ij}^0$ to different locations is a representation of social norms which are common rules in people's social life.

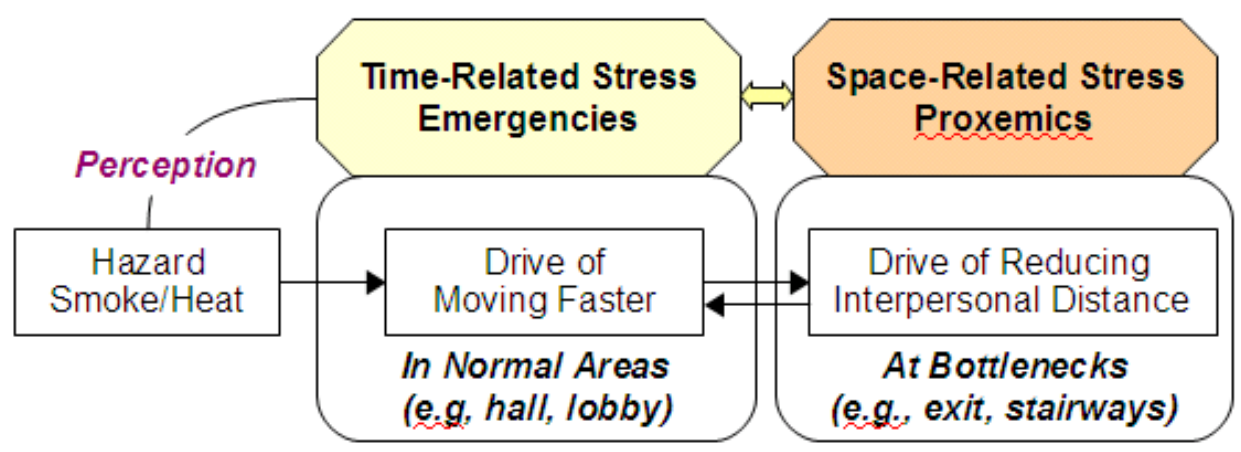

**Figure 1.4. Two Types of Stress Transforming Mutually: The sense of emergencies creates a kind of time-related stress which drives people to move fast in escape. At bottlenecks (e.g., narrow doorways) people cannot speed up as desired, and thus time-related stress is transformed into space-related interpersonal stress in order to pass through the bottleneck quickly.**

In emergencies the social norm is adjusted and competitive behavior may emerge, and the model is thus applied to simulation of jamming and stampede at narrow passageways. Such narrowings are commonly identified as bottlenecks for crowd egress. Especially, the sense of emergency initially produces time-related stress that motivates people to move fast. At bottlenecks such motion cannot be realized as desired, and the time-related stress is transferred to the space-related stress as shown in Figure 1.4. The desired interpersonal distance at bottlenecks is thus reduced, and people would like to "compress" their personal space in order to pass through the bottleneck quickly. This effect actually exhibits a kind of collective intelligence to increase transport efficiency at the bottleneck. The social norm is thus modified such that $d_{ij}^0$ is scaled down at bottlenecks. The parameter of $A_i$ may also be scaled down so that the social force as a whole is reduced in such an occasion (Korhonen, 2018).

To testify the above effect at bottlenecks we slightly modify the source program of FDS+Evac to implement the desired interpersonal distance (Hostikka, et. al., 2007; Korhonen, et. al., 2008; Korhonen, 2018). The social force is modified as below (Wang, 2016).

$$\boldsymbol{f}_{ij}^{soc} = A_i \exp\left[\frac{(d_{ij}^0 - d_{ij})}{B_i}\right]\boldsymbol{n}_{ij} \quad \text{or} \quad \boldsymbol{f}_{ij}^{soc} = \left(\lambda_i + (1-\lambda_i)\frac{1+\cos\varphi_{ij}}{2}\right) A_i \exp\left[\frac{(d_{ij}^0 - d_{ij})}{B_i}\right]\boldsymbol{n}_{ij} \qquad (1.3)$$

Here $A_i$ and $B_i$ are positive constants, which affect the strength and effective range of the repulsion between two individuals. The distance of individual $i$ and $j$ is denoted by $d_{ij}$, and $\boldsymbol{n}_{ij}$ is the normalized vector which points from individual $j$ to $i$. The gap of $d_{ij}^0$ and $d_{ij}$ implies the difference between the subjective wish in one's mind and objective feature in the reality, and it is an indication of interpersonal stress related to the social space, and it fits well in the social force $\boldsymbol{f}_{ij}^{soc} = \boldsymbol{F}^{soc}(d_{ij}^0 - d_{ij})$ in a general sense. The geometric features of two individuals are illustrated in Figure 1.5. Moreover, an anisotropic formula is widely used where Equation (1.3) is scaled by a function of $\lambda_i$. Here $\varphi_{ij}$ is the angle between $\boldsymbol{v}_i$ and $\boldsymbol{n}_{ji}$. If $\lambda_i = 1$, the social force is isotropic. $0 < \lambda_i < 1$ implies that the force is larger in front of an individual than behind.

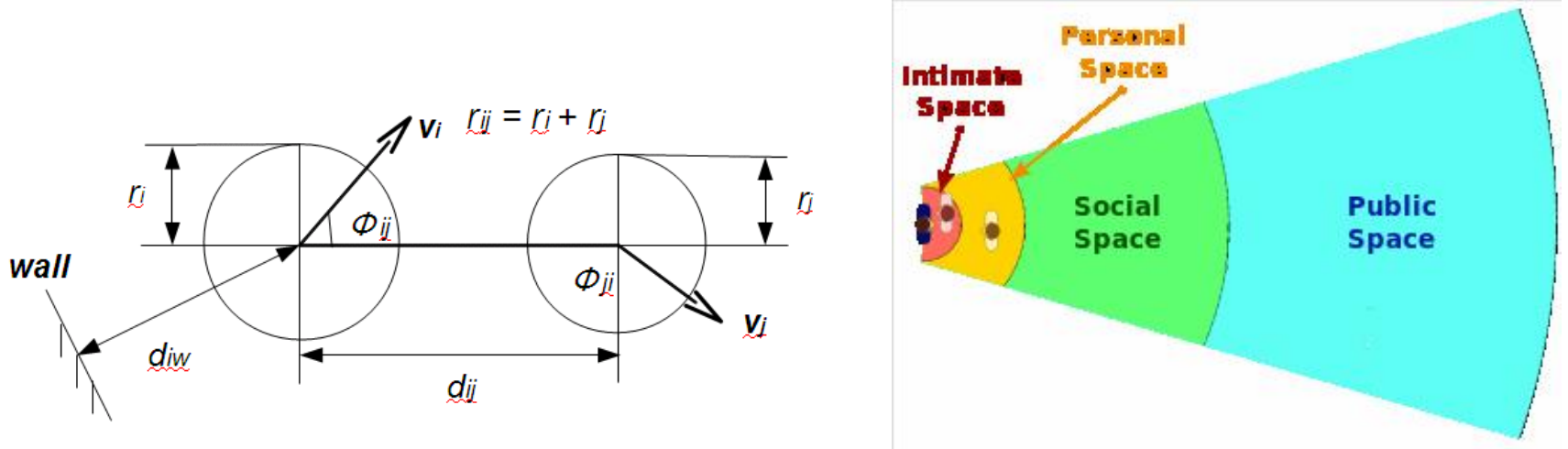

**Figure 1.5. A Schematic View of Two Individual Evacuees (See Equation 1.3). The social interaction and proximal distance origins from basic human instincts that we define a personal space to get a sense of safety. People normally feel stressed when their personal space is invaded by others. There are four interpersonal distances mentioned in Hall, 1966: intimate distance (<0.46m), personal distance (0.46m to 1.2m), social distance (1.2m to 3.7m), and public distance (>3.7m).**

Below is the simulation result by using FDS+Evac, and the example is based on IMO door flow test (IMO, 2007), where the door width is 1m, and it is also the door width used in Helbing et. al., 2002 and 2005. Let $r_{ij}$ denote the sum of the radii of individual i and j, namely, $r_{ij} = r_i + r_j$ The left diagram corresponds to large $d_{ij}^0$, where we specify $d_{ij}^0 = 3 \cdot r_{ij}$ , while the middle diagram corresponds to relatively small $d_{ij}^0$, where $d_{ij}^0 = 2 \cdot r_{ij}$ is used. The comparative results suggest that decreasing desired distance $d_{ij}^0$ moderately will increase the egress flow rate at the bottleneck. This result explains why people tend to reduce their interpersonal distance at the entrance or exit because such behavior increases the egress flow rate and thus reduce egress time.

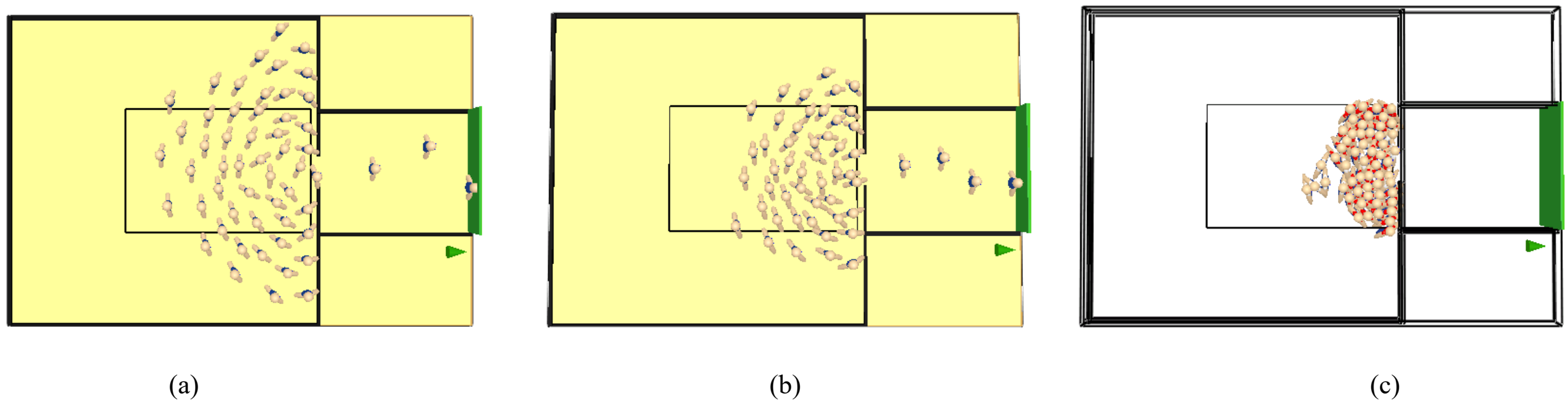

(a) (b) (c)

**Figure 1.6. Social Force and Blocking Effect: (a) Use large $d_{ij}^0$ in normal situation such that people obey social norm of large interpersonal distance. The result is decrease of flow rate and less chance of physical interaction. (b) Use small $d_{ij}^0$ in emergency egress such that people follow the social norm of small interpersonal distance. Flow rate thus increases and physical interaction increase in a stochastic sense. (c) As $d_{ij}^0$ continues to decrease, the physical interaction causes someone to fall down, and the doorway is thus clogged by those falling-down people. The comparative testing results are obtained by modifying the source code of FDS+Evac (Version 6.5.3), and the data can be downloaded at https://github.com/crowdmodel/test-crowd-dynamics**

By introducing desired distance $d_{ij}^0$ we suggest that two types of stressors could transform as shown in Figure 1.4. The emergencies creates a kind of time-pressure which motivates one to move fast. Such time-related stress is transformed to space-related stress at bottlenecks. In current version of FDS+Evac such transformation happens to be modeled by tuning parameter $A_i$ , where $A_i$ = 2000 max(0.5, $|\boldsymbol{v}_i|/|\boldsymbol{v}_i^0|$) is a function of $|\boldsymbol{v}_i|/|\boldsymbol{v}_i^0|$. This method reduces the strength of social force if $|\boldsymbol{v}_i|<|\boldsymbol{v}_i^0|$. As a result, people will get closer to each other if $|\boldsymbol{v}_i|<|\boldsymbol{v}_i^0|$, indicating that the time-related stress is transformed into a space-related issue. Here the ratio of actual speed and desired speed partly indicates time-related stress, and it is analogous to the difference $\boldsymbol{v}_i^0 - \boldsymbol{v}_i$ in term of only speed, and the direction of velocity is ignored. In a similar manner we bring another formula by tuning $d_{ij}^0$ as below $d_{ij}^0 = d_{ij}^{0\text{-preset}}$ max(0.2, $|\boldsymbol{v}_i|/|\boldsymbol{v}_i^0|$), where $d_{ij}^{0\text{-preset}}$ is the desired interpersonal distance in normal situation and 0.2 is the lower bound factor. As a result $d_{ij}^0$ is a function of $|\boldsymbol{v}_i|/|\boldsymbol{v}_i^0|$, and it decreases if $|\boldsymbol{v}_i|<|\boldsymbol{v}_i^0|$, implying that the time-related stress also impacts the desired distance $d_{ij}^0$ and the social force. As a result, the social force decreases at bottlenecks. However, the trade-off of reducing $d_{ij}^0$ at bottlenecks is increase of possibility of physical interaction among people, which is the major cause of crowd disasters such as stampede.

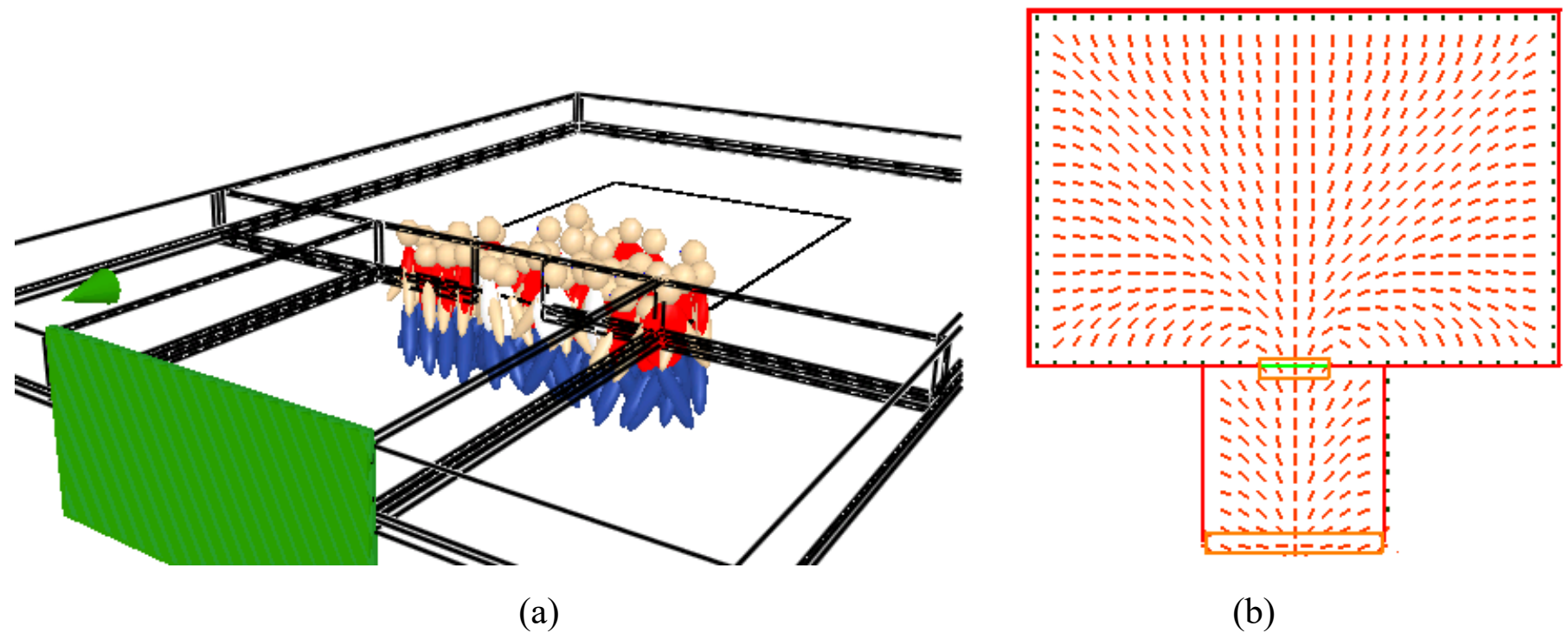

(a) (b)

**Figure 1.7. Crowd Escape at Bottleneck with Falling-Down Model: The white agents are falling down people who cannot move and are considered as obstacle to other agents. They fall down because the physical force exceeds a given threshold. The red agents are moving toward exit, and they have to get over the white ones to reach the door and the egress flow rate is thus significantly decreased.**

In fact scaling down $d_{ij}^0$ at bottleneck implies a sort of competitive behavior among people. In other words reducing $d_{ij}^0$ motivates people to reduce their physical distance $d_{ij}$, and this trend leads to high-density crowd, where the physical force might become effective and people may have more physical interaction. If physical force is intensified, someone may fall down. The falling-down people further impedes surrounding ones and slow down the egress flow significantly, and they may cause others to fall down again and stampede becomes a possible result. In sum the social force model with $d_{ij}^0$ is useful to study crowd behavior when jointly used with a falling-down model. As below FDS+Evac is used to simulate a falling-down event where evacuees fall down when the physical force exceeds a threshold (See Figure 1.6(c) and Figure 1.7).

In the above simulation $d_{ij}^0$ is manually changed to compare different scenarios. In fact $d_{ij}^0$ can be better adapted to a flow field as shown in Figure 1.7(b), where $d_{ij}^0$ is in inverse ratio of flow density. This setup may require a compressible fluid model where flow density varies with locations (Wang, 2016). In a psychological sense such a flow field can be better understood as a kind of social field (Helbing et. al., 2005; Lewin, 1951), which is used to adapt interpersonal distance in emergency egress.

## 2. Herding Effect and Pre-Movement Behavior

The phenomenon of herding widely exists in nature, not only for human crowd, but also among many socialized animals (e.g., a herd of sheep, a flock of birds). Herd members benefits from joining large groups, many of which refers to keep safety from outside environment such as predictors. In this section we will mainly discuss herding effect in emergency evacuation for human crowd, and an opinion dynamics is formulated as a linear model, and the model is applied in simulation of pre-movement behavior in crowd egress.

### (a) Herding Effect

Herding is especially evident when people are responding to an emergency (Low, 2000). Emergency implies time-pressure as mentioned before, and excessive time-pressure weakens the ability of logical thinking and reasoning, and independent decision making is more difficult in stressful conditions. Thus, people are more inclined to follow others (e.g., surrounding others' decision) rather than make decisions by themselves, and such herding behavior help people to gain a sense of safety when facing uncertainty. This is evident in social psychological study. Such herding effect is a rooted nature in many specifies of social animal such as in sheep herd, bird flock or fish school, and it is thus a base instinct for many individuals to keep safety against the environment with uncertainty. Based on opinion dynamics and social-force model, the computational model of herding behavior is generally given as below (Deffuant et. al., 2000).

$$opinion_i(t+1)=(1-p_i)\,opinion_i(t)+p_i[others(t)]_{\rightarrow i} \qquad [others(t)]_{\rightarrow i}=\sum_{R_i} c_{ij}\,opinion_j(t) \quad \sum_j c_{ij}=1 \tag{2.1}$$

where $opinion_i(t)$ represents opinion of individual agent $i$ at time point $t$, and it is a continuous real number or vector. Here $c_{ij}$ is non-zero if individual $i$ is able to perceive or acquire opinion of individual $j$ such that individual $i$ is thus influenced by individual $j$. A simple example is assuming that $c_{ij}$ becomes non-zero if the physical distance is less than $R_i$, namely, $d_{ij}<R_i$. In other words individual $i$ is able to know surrounding people's opinions within the range of $R_i$ by talking or observing their behavior. A more useful method is using the value of $c_{ij}$ as quantitative measurement of social relationship among individuals. Usually one's opinion is more affected by those who have close social relationship, and the value of $c_{ij}>0$ is larger for such individuals, such as family members or friends. By normalizing parameter of $c_{ij}$ with $\sum_j c_{ij}=1$, it is reasonable to define a matrix $C=[c_{ij}]_{nxn}$ for quantitative measurement of such social relationship. Such a matrix $C$ could be represented by a directed graph. Take three individuals for example (See Figure 2.1), where the directed arc from individual 1 to 2 means that individual 2's opinion is impacted by individual 1 with a normalized weight $c_{21}=0.7$. The initial values of $x_i(t)$ at $t=0$ are 13, 5 and 21 for agent $i=1,2,3$. The decision balance parameter $p_i$ is given by $p_1=0.3$, $p_2=0.5$, $p_3=0.25$. Their opinions converge in the simulation timeline as illustrated in Figure 2.1.

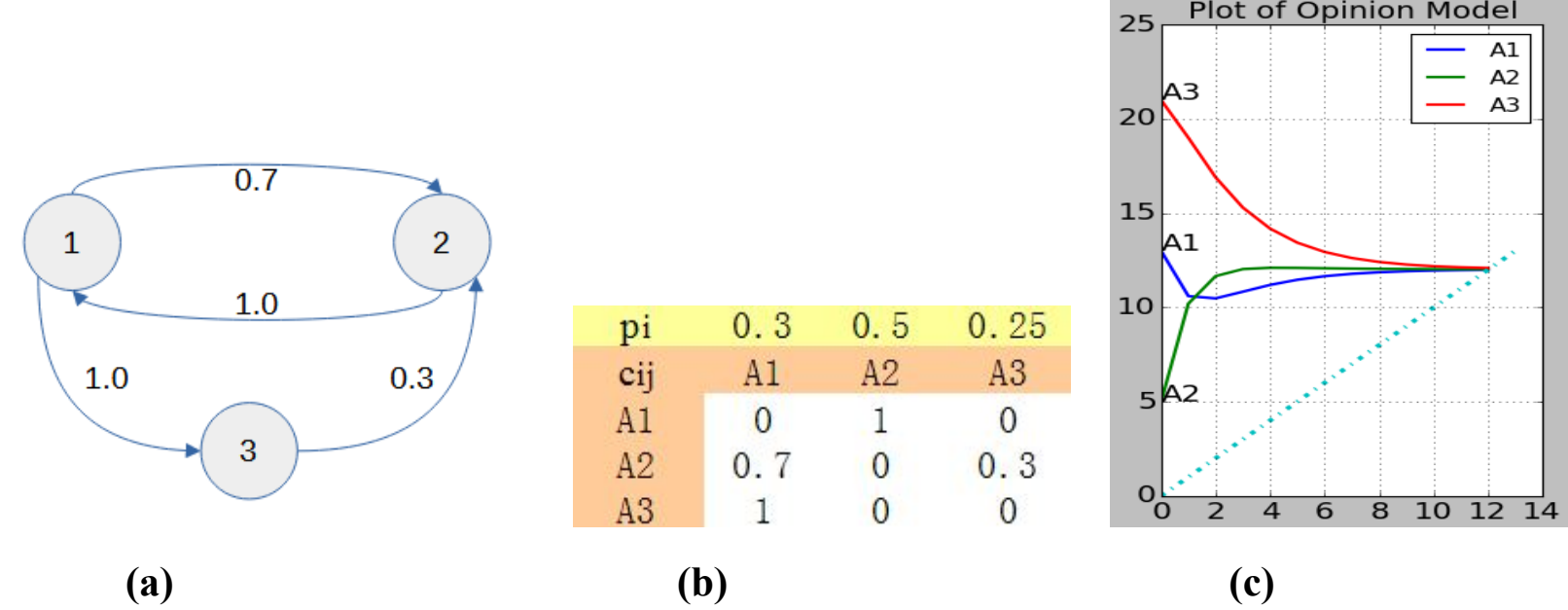

| pi | 0. 3 | 0. 5 | 0. 25 |
| --- | --- | --- | --- |
| cij | A1 | A2 | A3 |
| A1 | 0 | 1 | 0 |
| A2 | 0. 7 | 0 | 0. 3 |
| A3 | 1 | 0 | 0 |

**Figure 2.1. Social Topology and Opinion Dynamics: (a) The graphic connectivity of three agents; (b) The numerical valuation of matrices $\Pi = diag(p_1, p_2, \dots p_n)$ and $C=[c_{ij}]_{nxn}$ in consistency with the graphical model; (c) The numerical testing results of opinion dynamics by using Equation (2.1).**

An issue to be emphasized is on the diagonal element of $C=[c_{ij}]_{nxn}$. We define $c_{ii}=0$ if there exists any input arc to individual $i$ in the social topology. If not, we will let $c_{ii}=1$ to meet the requirement of $\sum_j c_{ij}=1$. Thus, for practical computing, Equation (2.1) is applicable to either an isolated individual or an individual who is socially connected to others. As for any isolated individual who has no input social connection from other individuals, it implies that $c_{ij}=0$ for $\forall j\neq i$, and we define $c_{ii}=1$ on the self-loop such that $opinion_i(t)$ is not changing with time. This definition looks a little intricate, but it is useful and consistent when we discuss time-variant topology because the changing $c_{ij}$ is not coupled with $p_i$, and we are thus able to discuss how $C=[c_{ij}]_{nxn}$ and $\Pi = diag(p_1, p_2, \ldots p_n)$ are dynamically changing as two separate issues.

Now it is feasible to further integrate parameter $p_i$ with matrix $C=[c_{ij}]_{nxn}$, describing how the opinion of individual $i$ is socially formed from matrix $W=[w_{ij}]_{nxn}$. Based on Equation (2.1), matrix W is given by $W = (I-\Pi) + \Pi\cdot C$, where $\Pi = diag(p_1, p_2, \ldots p_n)$ is a n×n square matrix with its diagonal element defined by $p_1, p_2, \ldots p_n$, and $I$ is a n×n the identity matrix. In other words, we have $w_{ij} = p_i \cdot c_{ij}$ if $i \neq j$. If $i=j$, we have $w_{ii} = (1-p_i) + p_i \cdot c_{ii}$. If there exists any input arc from individual $j$ to individual $i$ in the topology, it implies $c_{ii}=0$, and we have $w_{ii} = (1-p_i)$ for any individual with input social connection from others. As for an isolated individual, it implies $w_{ii} = (1-p_i)+ p_i \cdot c_{ii} =1$ because we define $c_{ii}=1$ for any isolated individual as mentioned before. It is easy to prove that $\Sigma_j w_{ij}=1$, meaning that $W=[w_{ij}]_{nxn}$ is a standard stochastic matrix, i.e., a non-negative matrix with all its rows summing up to 1.

The matrix $W=[w_{ij}]_{nxn}$ summarizes how individuals will keep his or her own opinion or follow others' opinion, and it enables us to formulate an opinion model as a linear dynamical system. Suppose the opinions of $n$ individuals are vertically stacked into a column vector $OPIN(t)= (opinion_1(t), opinion_2(t), \ldots opinion_n(t))^T$, and Equation (2.1) is compactly written by

$$OPIN(t+1)=W \cdot OPIN(t) \tag{2.2}$$

Here $OPIN(t)$ is the opinion profile of $n$ individuals at time $t$, and their opinions are updated in an iterative way based on Equation (2.2). If $W=[w_{ij}]_{nxn}$ is time-invariant, the opinion dynamics of Equation (2.2) is the classical DeGroot model (French 1956, DeGroot 1974, Lehrer 1975), which is theoretically similar to a homogeneous Markov process. The existing theory in linear algebra suggests that opinion $OPIN(t)$ converges if matrix $W$ satisfies certain conditions. For example, if matrix W is irreducible, the corresponding directed graph is strongly connected. With the classical case of constant weights a typical phenomenon is consensus (French 1956, DeGroot 1974, Lehrer 1975), and consensus of $n$ strongly-connected individuals means that $W=[w_{ij}]_{nxn}$ is primitive, namely, the power of matrix $W$ is strictly positive (i.e., does not contain any zero). Note that the above analysis is assuming simultaneous communication among agents by using Equation (2.2) in a given social topology. A more realistic computational method can be an asynchronous gossip-based algorithm, where only two agents are interacting in an asynchronous manner. The asynchronous algorithm for (2.2) is presented as below,

**Algorithm 1:** Pseudocode for asynchronous approach to (1):

**for** each simulation time t do

    **for** each agent $i$ in population do

        if agent $i$ is isolated: $x_i(t+1) = x_i(t)$

        else: randomly select an agent $j$ with $c_{ij}\neq 0$

            update opinion: $x_i(t+1)=(1- p_i c_{ij})x_i(t) + p_i c_{ij}\, x_j(t)$

        **end for**

**end for**

There is also an interesting variation of this model developed by Friedkin and Johnsen 1990, 1999. This model assumes that agent $i$ will always keep some prejudice to a certain extent $1 - g_i$ and by a susceptibility of $g_i$ the agent is socially connected with other agents.

$$opinion_i(t+1)= g_i \Sigma_j w_{ij}\cdot opinion_i(t) + (1-g_i)\, u_i(t) \qquad OPIN(t+1) = G\, W\cdot OPIN(t) + (I - G)\, u(t) \tag{2.3}$$

Here $G$ is also a diagonal matrix, namely, $G = diag(g_1 \;\; g_2 \;\ldots\; g_n)$ with $0 \leq g_i \leq 1$, and $I$ is the identity matrix. The prejudice $u_i(t)$ is a continuous real or vector which is in similar type to $opinion_i(t)$, and $OPIN(t)$ and $u(t)$ are thus vectors in the same dimension. It is often assumed that $u_i(t)$ is constant in the simulation process (Friedkin and Johnsen 1990, 1999). For example the initial opinion profile at $t=0$ is often used for a constant $u_i(t)=opinion(0)$.

In another sense $u(t)$ is an input signal in a dynamical process, and it could be considered as a noise signal which is randomly generated by certain probability distributions, describing a sort of random opinions as input signals, and the resulting system dynamics is given by $opinion_i(t+1)= \Sigma_j w_{ij}\cdot opinion_i(t) + \eta_i u_i(t)$, where $\eta_i \geq 0$ tunes the intensity of the random input signals. For example, we may assume $u_i(t)$ follow standard Gaussian distribution with with zero mean and unit standard derivation i.e., $u_i(t) \sim N(0,1)$, and $u_i(t)$ t=0,1,2… is a delta correlated scalar noise normally distributed in $(-\infty, +\infty)$. Such a random signal generates Gaussion white noise in the time sequence, and it describes randomness effect in one's opinion dynamics. From the perspective of psychology and behavior science, the random signal $u(t)$ can be interpreted as certain amount of irrational opinion in social interaction. As a result, the term of "panic" and irrational behavior are possibly modeled and simulated by integrating this noise signal $u(t)$ into opinion dynamics.

Now let us discuss a more complicated issue for time-varying $W$, where $w_{ij}(t) \geq 0$ is dynamically changing as time proceeds. This situation implies that either $C=[c_{ij}]_{nxn}$ or $\Pi = diag(p_1, p_2, \ldots p_n)$ is updated as agents move and interact. As for time-varying $C$, $c_{ij}(t) \geq 0$ is dynamically changing. This situation implies that the connectivity among agents is timely updated, and the social topology among agents thus become time-varying. In practical computing $p_i$ is updated relatively slower than $c_{ij}$ because $p_i$ reflects one's tendency to adhereing to his or her own opinion or following others, and it is an inherent characteristic in one's personality. In contrast $c_{ij}$ is more likely to be timely updated since it describes whether individual $i$ is able to acquire opinion of individual $j$, or in other words, whether individual $j$'s opinion has impact to individual $i$. For practical computing, it is feasible to further decompose $c_{ij}$ as two elements, namely $b_{ij}$ and $s_{ij}$ .

$$c_{ij}=\frac{b_{ij}s_{ij}}{\sum_k b_{ik}s_{ik}} \qquad b_{ij} = 1 \text{ if individual } i \text{ has access to acquire individual } j\text{'s opinion} \tag{2.4}$$

In Equation (2.4) $b_{ij}$ is a kind of Boolean variables, timely updated based on people's location and behavior, describing if people are able to exchange opinions (e.g., observing others' choice or talking with others to exchange opinions). In others words, Equation (2.1) only describes a class of opinion model, but it does not explicitly tell when it will be effective as individuals move in a two-dimensional space. Instead, Equation (2.5) initializes a discussion in what conditions one's opinion will be influenced by another. In existing theoretical models, there are two major branches relevant to the above modeling framework:

(i) Bounded Confidence Model:

The model suggests that interactions bring opinion closer to each other if they are already close sufficiently, and thus one's opinion is inclined to selectively follow similar opinions of others, and this algorithm will generate several small groups rather than one large group of consensus. Mathematically, the model assumes that each agent possesses a confidence range, and $b_{ij}(t)$=1 within the range (Hegselmann and Krause, 2002; Bernardo et. al., 2024). The model is opinion-dependent as a nonlinear dynamics.

$$b_{ij}(t)=1 \text{ if } |\, opinion_i(t)-opinion_j(t)\, | \leq R_{ij}; \qquad b_{ij}(t)=0 \text{ otherwise} \tag{2.5}$$

With Equation (2.1), (2.4) and (2.5), an interesting issue is that matrix $C$ and W cannot be mathematically expressed as an explicit function of time $t$, but an explicit function of states of $n$ individuals, and in return they also determine their state in the next time step. Consequently, the opinion dynamics of $n$ individuals becomes nonlinear in nature. As below we will briefly introduce some basic models which have been widely studied and used in existing research.

In the homogeneous bounded confidence model, the confidence ranges are the same for all the agents, i.e., $R_{ij}=R$ (Hegselmann and Krause, 2002). However, we must note that the heterogeneous model allows us to capture more realistic behavior in a social context, and the confidence ranges vary from individuals such that their interactions can be asymmetric. As a result, an agent are selectively influenced by another, and their interactions can be asymmetric. Existing research for heterogeneous model mainly exist in an exhaustive numerical analysis. From the perspective of social science, it is reasonable to assume that $R_{ij}$ and $c_{ij}$ should be dependent issues because both of them characterize the relationship among individuals. Thus, in a general sense, we may define $R_{ij}$ as a function of $c_{ij}$, and $R_{ij}$ is replaced by $R_{ij}=h(c_{ij})$. If $h(c_{ij})$ is a constant, i.e., $h(c_{ij})=R$, the model defined by Equation (2.1) and (2.4) is degenerated to the homogeneous model. Another common type is using a linear form of $h(c_{ij})$ such as $R_{ij}=Rc_{ij}$, where the baseline confidence range $R$ is multiplied by relational factor $c_{ij}$. The linear form $R_{ij}=Rc_{ij}$ ensures that $b_{ij}(t)=0$ if $c_{ij}=0$, and this feature is due to the interacting of any two agents is only enabled if $c_{ij}>0$. If $c_{ij}=0$, it implies that $R_{ij}=Rc_{ij}=0$, and it follows $b_{ij}(t)=0$, and thus the agent $i$ is not influenced by agent $j$.

In another sense we note that (2.1) slightly reshape the classical DeGroot Model for it decomposes the model by two sets of parameters $\Pi = diag(p_1, p_2, \ldots p_n)$ and $C=[c_{ij}]_{nxn}$ . This mathematical description enables us to study changing topology $C=[c_{ij}]_{nxn}$ and coefficient $\Pi$ as two separate issues. A special model is known as the Hegselmann-Krause model (HK model), where $\Pi$ and $C$ are both time-varying with homogeneous confidence range, i.e., $R_i=R$. In the HK model, the connectivity among agents is defined dynamically by $b_{ij}(t)$, and relational factor $c_{ij}$ is evenly valued as a constant for all the surrounding agents in connection, e.g., $s_{ij}$ =1/(n-1) for ∀i≠j. The herding coefficient $\Pi = diag(p_1, p_2, \ldots p_n)$ is also time-varying, depending on the connectivity states. The time-varying law for $\Pi = diag(p_1, p_2, \ldots p_n)$ is thus given by $1-p_i(t)=1/(1+\Sigma_j b_{ij}(t))$, and thus we have $p_i(t)=(\Sigma_j b_{ij}(t))/(1+\Sigma_j b_{ij}(t))$, which means that $p_i(t)$ is a function of $\Sigma_j b_{ij}(t)$, i.e., $p_i(t)$ depends on the connectivity matrix $[b_{ij}(t)]_{nxn}$ .

The remarkable merit of the classical HK model exists in the convergence in finite time steps to a constant steady state. This is different from using the constant $\Pi = diag(p_1, p_2, \ldots p_n)$ as static self-influence weight, which only ensures asymptotic convergence. Figure 2.2 illustrates the difference by specifying $p_i$ =0.5, $R$=3 with a constant relational factor $s_{ij}$ ∀i≠j for 20 agents. The initial opinion profile is given by $x_i(0)=i+2$. In comparison, the corresponding classical HK model is plotted with the same initial opinion profile, but $p_i$ is dynamically updated by $p_i(t)=(\Sigma_j b_{ij}(t))/(1+\Sigma_j b_{ij}(t))$. In other words, Figure 2.2(a) and Figure 2.2(b) only differ in whether using time-varying coefficients $p_i$ , and it shows that the HK model converges faster than the case using constant $p_i$ =0.5, which means that the time-varying law of $p_i(t)=(\Sigma_j b_{ij}(t))/(1+\Sigma_j b_{ij}(t))$ is effective to speed up convergence.

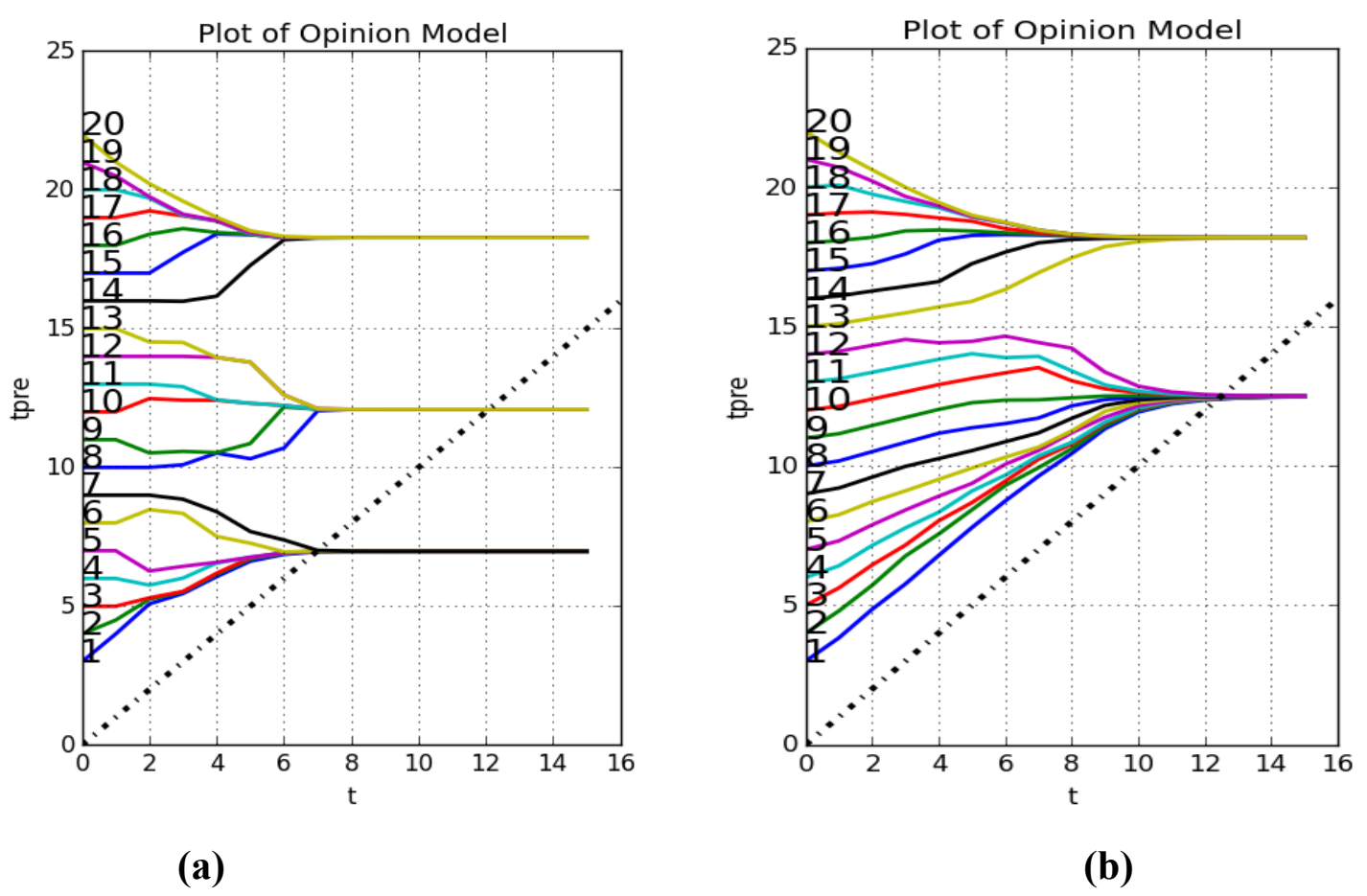

(a) (b)

**Figure 2.2. Bounded Confidence Models With and Without Using Time-Varying Law for $\Pi = diag(p_1, p_2, \ldots p_n)$. By comparing subplot (a) and (b), it well illustrates that the time-varying law of $p_i(t)=(\Sigma_j b_{ij}(t))/(1+\Sigma_j b_{ij}(t))$ is effective to speed up convergence.**

<u>(ii) Vicsek Model:</u>

The model assumes that this effect takes place only if the physical distance is less than $R_i$ , namely, $d_{ij}<R_i$ . Especially, the Vicsek flocking model assumes that each individual agent is given the same radius $R_i=R$, and thus the opinion of two agents are influenced in a mutual way. This is a reasonable and relatively straightforward assumption, and it is mainly obtained by observing a group of social animals (e.g., pigeons) in collective motion (Vicsek et. al., 1995), and such group members are relatively close to each other.

The Vicsek flocking model assumes spatial proximity to determine $b_{ij}$, and $b_{ij}$=1 if the spatial distance of two agents $d_{ij}(t)$ is within an interaction range $R$. In other words, if agent $j$ move into a circle of radius $R$ surrounding the agent $i$, the opinion of agent $i$ is influenced by agent $j$.

$$b_{ij}(t) = 1 \text{ if } d_{ij}(t)< R \qquad\qquad b_{ij}(t) = 0 \text{ otherwise} \tag{2.6}$$

The above modeling framework describes herding effect of many individuals, where $p_i$, $c_{ij}$ and $s_{ij}$ in Equation (2.1) and (2.4) characterize how agents selectively follow others based on their relational ties, and $b_{ij}(t)$ in Equation (2.6) specifies the effective range for herding behavior. The resulting multi-agent system is nonlinear and non-equilibrium, and it is often called many-particle simulation in the sense of statistical physics (Ginellia, 2016).

Based on existing model in opinion dynamics, there are some models that inspire us to modify this assumption for human crowd. For example, the social relationship of individuals is also of significant importance, and the interaction range is not only determined by physical distance $d_{ij}$, but also refer to social relationship of individuals, which is partly measured by the desired distance $d_{ij}^0$. As a result, one tends to follow those in close social relationship, but not only those in close positions. In existing literature various criteria are discussed to extend the basic rule of $d_{ij}<R$, and the resulting system can be much more complex than one would intuitively expect. We will further discuss this important issue in the following section of crowd group dynamics.

In our simulation platform crowdEgress (Wang et. al., 2023) there are several lists attached to each individual, including see list and attention list, which describe whether individuals could see each other or whether an individual pays attention to others. Such lists facilitates timely update of the value of $b_{ij}$ . Based on $b_{ij}$ we have a time-varying a directed graph. As below we illustrate such a time-varying graph as individual agents interact in social context, and this algorithm is being tested in our simulation platform crowdEgress (Wang et. al., 2023).

Now let us go back to the baseline model of (2.1) and (2.2). In brief the social relationship of $n$ individuals is described by $n$-dimensional matrix $C$ and matrix $W$. Here we can replace such opinions by desired velocity or desired interpersonal distance, and an example is given as below by replacing opinions by desired velocity.

$$v_i^0(t+1)=(1-p_i)v_i^0(t)+p_i[v_{others}^0(t)]_{\to i} \qquad [v_{others}^0(t)]_{\to i}=\Sigma_{R_i} c_{ij} v_j^0(t) \tag{2.7}$$

Based on Equation (2.7) the desired velocity $\boldsymbol{v}^0_i$ is updated by mixing itself with the average velocities of others within radius $R_i$. Both options are weighted with some parameter ($1-p_i$) and $p_i$ , and two options follow two-point distribution with probability ($1-p_i$) and $p_i$ , and $v^0_i$ is updated by the weighted average. As a consequence, individualistic behavior is dominant if $p_i$ is low whereas herding behavior increases if $p_i$ goes increasing.

In principle parameter $p_i$ indicates how an individual keeps balance between his or her own opinion and opinions of others, namely, whether an individual is stubborn or open-minded. One model refers to how $p_i$ is dynamically adjusted through the simulation process. For example, in Helbing, Farkas and Vicsek, 2000 and Helbing et al., 2002, $p_i$ is considered to indicate one's panic level, which is given by ratio of $(v_i^0 - v_i)/v_i^0$, and is called a “nervousness" parameter. This ratio critically affects several testing results in their work. As mentioned before the gap of $(v_i^0 - v_i)$ is understood as an indicator of one's stress level (See Table 1.1), and it is further normalized by dividing $(v_i^0 - v_i)$ by $v_i^0$. As a result, the “nervousness” parameter $(v_i^0 - v_i)/v_i^0$ can be explained as a normalized stress indicator, and it shows that people are more inclined to follow others when they feel more stressed in an emergency situation. However, it not suitable to directly use $(v_i^0 - v_i)/v_i^0$ to calculate desired velocity based on Equation (2.7) because $(1-p_i)v_i^0$ is simply $v_i$ if we assume $p_i=(v_i^0 - v_i)/v_i^0$ . In fact the stress perceived for us is not instantaneous, but accumulating with a period of time, and it is necessary to model such accumulating stress in timeline. Thus, a general formula to update parameter $p_i$ is given by a function such as $p_i(t) = h(t, \boldsymbol{v}_i^0 - \boldsymbol{v}_i, d_{ij}^0 - d_{ij})$. As a result, parameter $p_i$ is directly an indicator of one’s stress level as shown in Table 1.1. In sum how to compute parameter $p_i$ in a dynamical process is an interesting study topic and we will further discuss this issue in the future.

### (b) Pre-Movement Behavior

In brief the evacuation behavior in building egress is commonly divided into two phase: the pre-movement phase and movement phase. In the pre-movement phase evacuees try to interpret the alarm signal received. They normally collect and exchange information, searching for new cues to confirm what really happens instead of directly moving to an exit or safe place. Existing egress research suggests that pre-movement behavior critically affects the total egress time in building evacuation, and it further consists of recognition and response to environmental stressors as shown in Figure 2.3. The recognition time refers to the time interval from the first cues detected by an evacuee (i.e. fire alarm, warnings, etc.) to the time of awareness that one need to respond to the new situation. During this time interval, evacuees normally seek further information to interpret the perceived cues and try to find out what exactly happens. The response time is the time interval from awareness of potential threats to the time to start leaving for a safety place. Different actions can be performed during this interval such as calling for firefighters, collecting personal belongings or socially connecting with others. According to fight-or-flight response (Cannon, 1932; Adam and Gaudou, 2017), leaving is one option, but seldom the first option.

Very importantly, if many individuals are involved in evacuation, leaving for safety is rarely an individualistic action, but a group decision in a collective sense. In other words, people are inclined to be socially tied with others to respond to the situation as a group, and such social attachment behavior helps to relieve stress, providing group members with a sense of safety. However, the trade-off is that it often takes certain time for people to form a social group, and opinions of the group members may vary to some extent such that it also takes time for them to reach an agreement. Here the group size and social relationship of the group members are among the key factors to determine the recognition and response time (Sorensen, 1991), and things often become better if there is an effective leader in the group, for example in occasions of office buildings or schools. The reason is that there is commonly no responsibility diffused or inhibited in these place, and it becomes easier to form the leader-and-follower pattern in collective behavior.

In the past 30 years various behavioral models were proposed in the circles of psychological and behavior science (Sime, 2001; Kuligowski, 2009). These models are usually simplified in the engineering field for practical use in computation, resulting in state-transition model (Lovreglio, Ronchi and Nilsson, 2015; Adam and Gaudou, 2017) or sequential flow to describe human response to environmental stimuli (e.g, fire alarm). When evacuees confirm the potential risk in the environment, their states transit from the pre-movement phase to the movement phase and they start to move to an exit for safety.

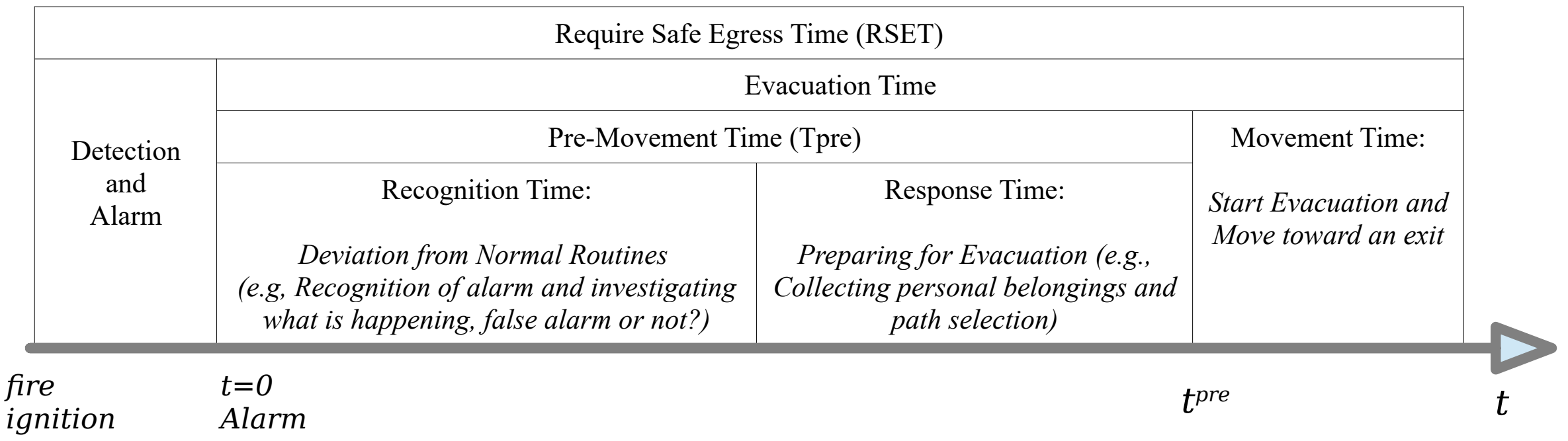

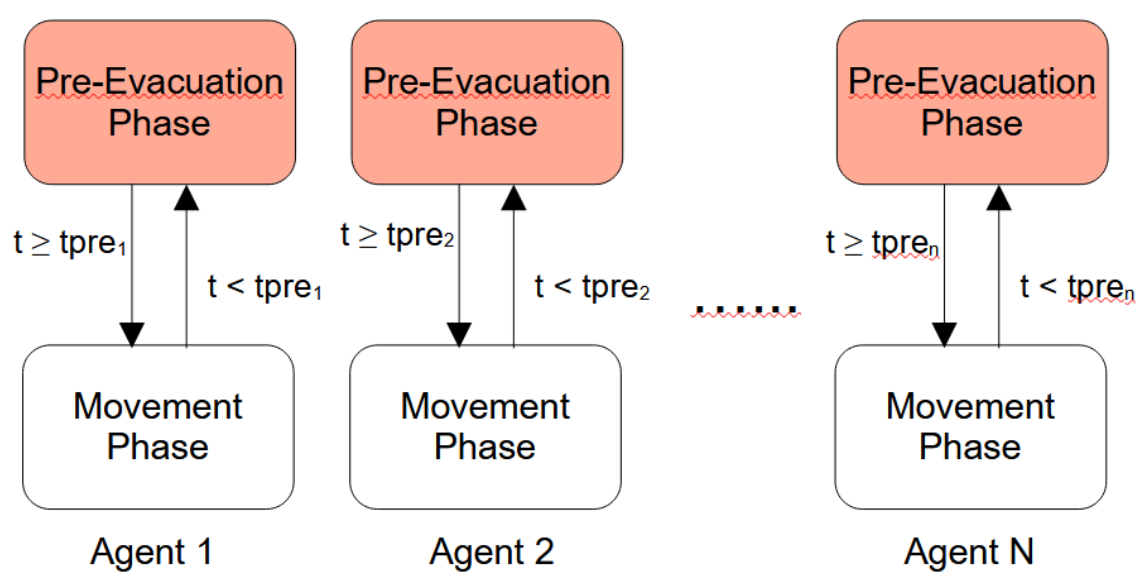

**Figure 2.3. Transition from Pre-movement Phase to Movement Phase: Such phase transition takes place if simulation time *t* exceeds the an individual's pre-movement *time* $tpre_i$ . Each individual's $tpre_i$ is initially different, and social interaction bring $tpre_i$ to converge in certain conditions.**

In this article we will integrate a simple version of phase transition model with opinion dynamics to calculate the pre-movement time of many individuals, and this model mainly describes social interaction in this phase, facilitating to investigate how collective pre-movement time emerges from individual interactions in a given social topology. The model critically captures the social structure of many individuals including the group size and the possible leader-and-follower effect, and the computational method has been tested in our simulation platform crowdEgress (Wang et. al., 2023) as well as FDS+Evac (Korhonen, 2018).

In the following discussion the term "agent" is often used to represent evacuees. Let $n$ be the total number of agents under consideration. All the agent are sequentially indexed by $i$=0,1, ... $n$-1. We assume that simulation starts at t=0, and the pre-evacuation time for $n$ evacuee agents are initialized as a continuous real number $tpre_i$(t)>0 at t=0. Conceptually, $tpre_i$(t) measures the psychological state of evacuees, and it refers to perception of time-pressure or subjective feeling of urgency, which is caused by the outside environmental stressors (e.g., fire alarm). Such perception is individualized because different people may perceive the same situation differently.

Because $tpre_i$(t) reflects individualized perception of time-pressure, it should also fit in physics scope of measurement of time. In real-world evacuation scenarios $tpre_i$(t) could be in unit of seconds, minutes or hours, depending on different types of emergency events. Given $t$ as the physics time, also in unit of seconds, minutes or hours, and $tpre_i$(t) for $t$=0, 1, 2… reflects how agent $i$ feels the level of urgency in simulation timeline. At each time stage $t$=0, 1, 2…, the agents' pre-movement time is computed by interactive opinion dynamics based on Equation (2.1).

$$tpre_i(t+1)=(1-p_i)tpre_i(t)+p_i tpre_{others}(t) \qquad tpre_{others}(t)=\sum_j c_{ij} tpre_j(t) \quad \sum_j c_{ij}=1 \tag{2.8}$$

The pre-movement time of an individual $tpre_i$ is updated by mixing itself with a weighted average of others in certain social relationship. Both options are weighted with some parameter (1-$p_i$) and $p_i$ , and $tpre_i$ is updated by the weighted average. Here $c_{ij}$ is non-zero if individual $i$ has access to acquire individual $j$'s pre-movement time $tpre_j$. As mentioned before, matrix $C=[c_{ij}]_{nxn}$ defines how individuals are socially related to each other and whether they are able to perceive others' opinion in collective motion. The stochastic matrix $W$ in Equation (2.2) is deduced with its elements $w_{ij} = p_i c_{ij}$ if $i$ is not equal to $j$. If $i$=$j$, we have $w_{ii}$ = (1-$p_i$)+ $p_i \cdot c_{ii}$ such that $p_{ii}$=(1-$p_i$) for an individual with input social arc from others and $p_{ii}$=(1-$p_i$)+$p_i$ $c_{ii}$ =1 for an isolated individual.

From the psychological perspective of stress theory, the above binary phase transition symbolizes the fight-or-flight response, in which individuals perceive the threat and decide in a binary manner to react to a potential or actual threat [1], and it is the major underlying cause of pre-evacuation time period, explaining why evacuees often search for cues by instinct to confirm the situation instead of directly retreating. This mental process can be marked by physical adaptive changes such as nervous and endocrine changes that prepare an individual to either defend or retreat. The functions of this response were initially described in the early 1900s by American neurologist and physiologist Walter Brandford Cannon, who also coined the phrase (Cannon, 1932).

The above binary state machine could be extended by several methods. For instance, we may assume that each agent maintains two time parameters including the recognition time ($trec_i$) and response time ($tres_i$), and it yields $tpre_i = trec_i + tres_i$ . Accordingly, each automaton consists of three states where the recognition state goes to response state if $t > trec_i$ and response state further goes to movement state when $t > trec_i + tres_i$ . The opinion dynamics formulated by Equation (2.1) is applied to both $trec_i$ and $tres_i$. However, the things become a little complicated if agents in different states interact, for example, an agent in response state meet another agent in recognition state, and certain logic models should be involved. This extended model is more consistent with existing psychological findings in evacuation research. In this article, for simplicity, we will only discuss the binary state transition model as illustrated in Figure 2.3.

In contrast $s_{ij}$ defines a quantitative measure of social relationship, and it is a real number which weighs how much one's opinion is possibly impacted by others. Such social relationship is relatively stable, and it is not easily changed compared with parameter $p_i$ . As for opinion model of pre-movement time, it may also be feasible to extend the effective range of $s_{ij}$ such as $s_{ij} \in [-1, 1]$. As a result, $s_{ij}$ is not a probability measure, but generally a weight parameter which causes the individuals' opinions to either converge or diverge. $0< s_{ij} \leq 1$ means that individual $i$ intends to stay in somewhere between his or her own opinion and opinion of individual $j$, and this moderate strategy often brings opinions to converge in a sense. In contrast, $-1\leq s_{ij} <0$ means that one is against others' opinion. In other words, the negative value of $s_{ij}$ implies disagreement such that others' standpoint has an inverse effect on one's opinion. As a result, the more individual $i$ acquires opinion of individual $j$, the more he or she will reject it and hold more firmly on his or her own standpoint. This strategy brings the individuals' opinions not to converge, but diverge to some extent. In sum, $s_{ij} \in [-1, 1]$ implies a more complex social connection among individuals, where interactions bring opinions either closer to each other, or more apart from each other as shown in Table 2.1. This model is complex in theoretical analysis, but it is more comparable to real-world situations because people do not always agree with each other. Disagreement exists in many scenarios. For example, it has been reported in many bushfire events that the younger generation would like to follow the public instructions and leave their house for safety, but the elderly family members might insist staying at the house (Adam and Gaudou , 2017). Such disagreement will significantly delay response to fire alarm in a collective sense.

Table 2.1. Range of Parameter $p$ in Opinion Dynamics

| $-1\leq s_{ij}<0$ | $s_{ij}=0$ | $0<s_{ij}\leq 1$ |
|---|---|---|
| Against others' opinions | Hold his own opinion and do not care about others' opinions | Support others' opinions |

If we consider the negative value of $s_{ij}$ and extend the domain $s_{ij} \in [-1, 1]$, Equation (2.5) and (2.8) should be modified as below such that domain of $c_{ij}$ is also extended correspondingly as $c_{ij} \in [-1, 1]$. It is not difficult to find that $C^*=[|c_{ij}|]_{nxn}$ is still a stochastic matrix, but $C=[c_{ij}]_{nxn}$ is not.

$$tpre_i(t+1)=(1-p_i\sum_j c_{ij})tpre_i(t)+p_i\sum_j c_{ij}tpre_j(t) \quad \sum_j |c_{ij}|=1 \qquad c_{ij}=\frac{b_{ij}s_{ij}}{\sum_k b_{ik}|s_{ik}|}$$

$b_{ij}= 1$ if individual $i$ has access to acquire individual $j$'s opinion

$b_{ij}= 0$ if individual $i$ has no access to acquire individual $j$'s opinion (2.9)

It is obvious that Equation (2.9) will be degenerated to Equation (2.5) if all the $s_{ij}$ are non-negative and thus $\sum_j c_{ij}=1$ holds, implying that $p_i\sum_j c_{ij}= p_i$ . If not, $\sum_j|c_{ij}|=1$ holds instead of $\sum_j c_{ij}=1$. In fact there is no need to confine $s_{ij} \in [-1, 1]$, and $s_{ij}$ is conceptually a relative measurement for the social relationship among individuals. As long as $s_{ij}$ is finite, (2.9) automatically ensures that $\sum_j |c_{ij}|=1$. Further, the asynchronous algorithm could be applied, which is widely called the randomized gossip algorithm, and the probability of interacting two agents is specified by $C^*=[|c_{ij}|]_{nxn}$ .

By using Equation (2.9) opinions do not converge between any antagonistic agents, and certain gap exists in opinions of them. In a similar manner we have matrix $W=[w_{ij}]_{nxn}$ deduced from Equation (2.9), with its elements $w_{ij} = p_i c_{ij}$ if $i$ is not equal to $j$. If $i=j$, we have $w_{ii} = (1-p_i\sum_j c_{ij})+ p_i \cdot c_{ii}$ such that $w_{ii}=(1-p_i\sum_j c_{ij})$ for an individual with input social arc from others and $w_{ii}=(1-p_i c_{ii})+p_i c_{ii}=1$ for an isolated individual. Here $W=[w_{ij}]_{nxn}$ is not a stochastic matrix because $w_{ij}$ can be negative, but $\sum_j w_{ij}=1$ still holds for all its rows summing up to 1.

For simple demonstration, we apply the bounded confidence model to the above opinion dynamics. The bounded confidence model suggests that the binary variable $b_{ij}(t)$ for $\forall j \neq i$ is determined by combination of agent opinions. For example, agents are socially connected if they share similar opinions in a confidence range $R$, and thus $b_{ij}(t)=1$ if the difference of opinion is within a given confidence range, i.e., $|tpre_i(t)-tpre_j(t)|<R$. The above model formulates a nonlinear process, describing how individual-level opinions interact in either agreement or disagreement. It is not difficult to prove that the above system dynamics is always marginal stable, i.e., $tpre_i(t)$ will be finite for $\forall$i.

In the following discussion we present a study case of the bounded confidence model, where the connectivity criterion $b_{ij}(t)=1$ if $|\,opinion_i(t)-opinion_j(t)\,| < 3$ (min). The connectivity of a time-varying social topology $[b_{ij}]_{nxn}$ is thus determined by criterion of $R_{ij}=R<3$ (*min*) as given in Equation (2.6). The social relationship of 6 agents are summarized in Table 2.2, and the agents are identified by their index number i=0,1,2...6. The $\Pi = diag(p_0, p_1, \ldots p_6)$ is indicated as the diagonal element in Table 2.2, and it indicates how each agent keeps balance between its own opinion and others' opinion. They are sequentially given by 0.3, 0.5, 0.25, 0.3, 0.5, 0.25. As $p_i$ increases, the corresponding agent *i* becomes more and more open-mind to others' opinions. Parameter $s_{ij}$ is specified as the other non-diagonal elements in Table 2.2, and it indicates how the agents are socially connected with each other. Here $s_{ij}$ are normalized within the range of [-1, 1]. In this table we can generally identify a group which consists of individual 1, 2, 3, and another group of 4, 5, 6. The value of $s_{ij}$ are positive inside the group while negative between the two groups. The initial pre-evacuation time is sequentially given by 7, 5, 1, 10, 9, 2 (minutes).

Table 2.2. Social Relationship Matrix $S=[s_{ij}]$ and Parameter $p_i$

| $p_i$ | *Agent1* | *Agent2* | *Agent3* | *Agent4* | *Agent5* | *Agent6* |
|---|---|---|---|---|---|---|
| *Agent1* | *0.3* | *0.3* | *0.6* | *-0.1* | *0* | *0* |
| *Agent2* | *0.6* | *0.5* | *0.3* | *0* | *0* | *-0.1* |
| *Agent3* | *0.2* | *0.5* | *0.25* | *0* | *-0.3* | *0* |
| *Agent4* | *-0.1* | *-0.2* | *0* | *0.3* | *0.1* | *0.6* |
| *Agent5* | *0* | *-0.2* | *-0.1* | *0.1* | *0.5* | *0.6* |
| *Agent6* | *0* | *0* | *-0.2* | *0.1* | *0.7* | *0.25* |

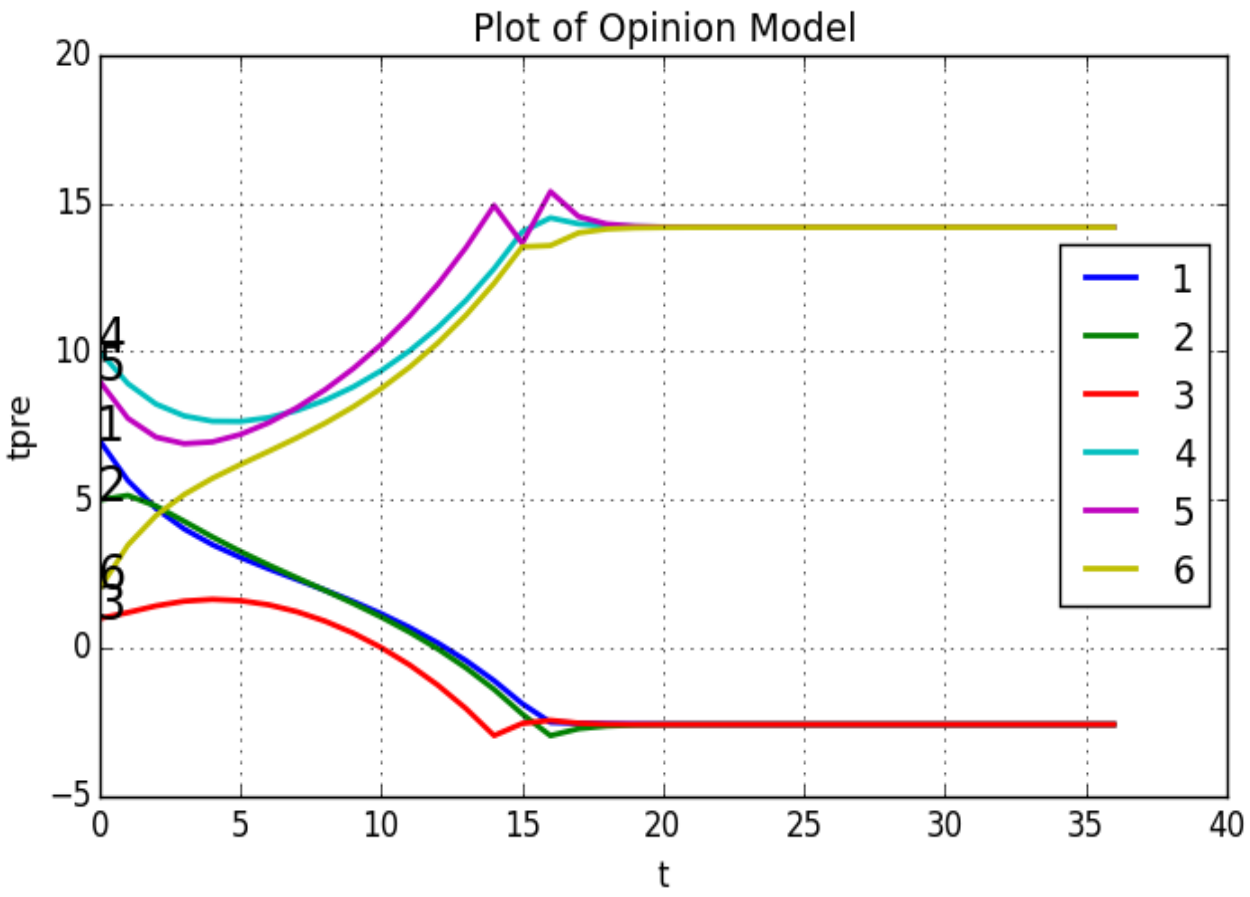

**Figure 2.4. Opinion Dynamics of Pre-Movement Time: The biparties effect is well illustrated in the above plot, where two groups are identified. The first group consists of individual 1, 2, 3 and the second group individual 4, 5, 6. The intra-group dynamics only include cohesion while the inter-group interaction include repulsion. The testing case could be founded and downloaded at https://sourceforge.net/p/socialnetwork-abs.**

In Figure 2.4 each colorful line represents $tpre_i(t)$ for each evacuee and the it dynamically changes as evacuees interact with each other. The figure is plotted by using our simulation platform, and the simulation results show the typical bi-groups phenomenon. The first group is composed by 1, 2, 3, and their opinions converge to about 4 (min), while the second group consists of 4, 5, 6, and their opinion converge to around 20 (min). In Figure 2.3 a dash line divides the plane equally into two regions. The upper left region is the pre-movement phase of $tpre_i>t$, and the lower right region is the movement phase for $t \geq tpre_i$ . When the colorful lines meet the dash line, state transition happens from the pre-movement phase to the movement phase.

Last but not least, the asynchronous version of a bounded confidence model is known as the Deffuant–Weisbuch (DW) model. At each time step, an agent selects a dyad (i.e., a pair of adjacent agents in the time-varying social network), and the two agents in the dyad interact. The algorithm is given as below.

**Algorithm 2:** Pseudocode for asynchronous approach:

**for** each time step **do**

Compute $b_{ij}(t)$ by Equation (2.6) or Equation (2.9) for $n$ agents.

Randomly select an arc in $[b_{ij}]_{n\times n}$ with $b_{ij}=1$

update opinion: $opinion_i(t+1) = (1-p_i c_{ij})\, opinion_i(t) + p_i c_{ij}\, opinion_j(t)$

$opinion_j(t+1) = (1-p_j c_{ji})\, opinion_j(t) + p_j c_{ji}\, opinion_i(t)$

**end for**

## 3. Social Group Behavior

People create group-level behavior beyond the ken of any single person, and in the past 20 years there has been growing realization in social science that such group-level organizations sometimes emerge spontaneously without any central design. Thus, it is reasonable to study such group phenomena in a bottom-up rather than a top-down manner. This approach creates computational units of individuals and their interactions, and to observe how the global structures are formed dynamically with their interactions. In evacuation process such group-level behavior widely exist in both pre-movement phase and movement-phase, and it significantly affects how people respond to environmental stressors such as fire alarms. For example, in emergencies people usually first seek for familiar ones (e.g., friends or parents) to exhcange information and then respond in a collective sense. Thus, it is meaningful to develop a social group model to characterize how an evacuee is affiliated with other familiar and trust ones, especially based on their social relationship. In a psychological sense, social group behavior agrees with the flight-or-affiliation effect (Bañgate et al., 2017), which is different from the fight-or-flight response (Cannon, 1932) as presented in Section 2. In this section we will elaborate this issue and introduce a force-based method to describe such social groups.

### (a) Group Social Force

In a group individuals exhibit some degree of social cohesion based on their relationship and they are more than a simple collection or aggregate of individuals. To model group dynamics, attraction is necessarily taken into account in interaction of individuals. For example, attraction will make acquainted people to join together and possibly form a group. In Helbing and Molnar, 1995 and Helbing et al., 2002 attraction was considered, but separate from the social force. In this paper attraction and repulsion are put in the same social context: repulsion makes people to keep proper distance while attraction makes them cohesive and form social groups. Thus, this subsection integrates attraction into the social force based on the concept of desired interpersonal distance. The resulting force is either repulsive or attractive, and it is especially useful to model group behavior in pedestrian crowd. The group social force is defined as below.

$$\boldsymbol{f}_{ij}^{soc} = \frac{A_{ij}}{B_{ij}}(d_{ij}^0 - d_{ij})\exp\left[\frac{(d_{ij}^0 - d_{ij})}{B_{ij}}\right]\boldsymbol{n}_{ij} \quad \text{or} \quad \boldsymbol{f}_{ij}^{soc} = \left(\lambda_i + (1-\lambda_i)\frac{1+\cos\varphi_{ij}}{2}\right)\frac{A_{ij}}{B_{ij}}(d_{ij}^0 - d_{ij})\exp\left[\frac{(d_{ij}^0 - d_{ij})}{B_{ij}}\right]\boldsymbol{n}_{ij} \qquad (3.1)$$

Here $A_{ij}$ and $B_{ij}$ are similar parameters as introduced before, and $\boldsymbol{n}_{ij}$ is the normalized vector which points from individual $j$ to $i$. The group social force also functions in a feedback manner to make the realistic distance $d_{ij}$ approaching towards the desired distance $d_{ij}^0$. A difference is that $\boldsymbol{v}_i^0$ and $\boldsymbol{v}_i$ are vectors while $d_{ij}^0$ and $d_{ij}$ are scalars.

When $d_{ij}$ is sufficiently large, the group social force tends to be zero so that individual $i$ and $j$ have almost no interaction. This trend is the same as the traditional social force (Helbing and Molnar, 1995 and Helbing et al., 2002). If $d_{ij}$ is comparable to $d_{ij}^0$ , interaction of individual $i$ and $j$ comes into existence. If $d_{ij}^0 < d_{ij}$, the group social force is attraction whereas it is repulsion if $d_{ij}^0 > d_{ij}$ . The attraction reaches the extreme value when $d_{ij} = d_{ij}^0 + B_{ij}$ , and the extreme value is $\boldsymbol{f}_{ij}^{soc} = -A_{ij}\exp(-1)$. The desired distance $d_{ij}^0$ makes the curve move horizontally with a certain interval, and it is the equilibrium position when an individual interacts with another one in pair. The curve shape is affected by parameter $A_{ij}$ and $B_{ij}$ . $A_{ij}$ is a linear scaling factor which affects the strength of the force whereas $B_{ij}$ determines the effective range of the interaction.

Two plots of Equation (3.1) are illustrated as above: Figure 3.1(a) shows that individual i is attracted by individual j when they are sufficiently close, and this suggests that individual i and j are probably familiar with each other. Figure 3.1(b) does not show such relationship because their interaction range and magnitude both reduce remarkably. In the above curve the negative segment represents attraction, and it represents a kind of social cohesion which facilitates to form groups. In contrast the positive segment denotes repulsion and it functions like the traditional social force.

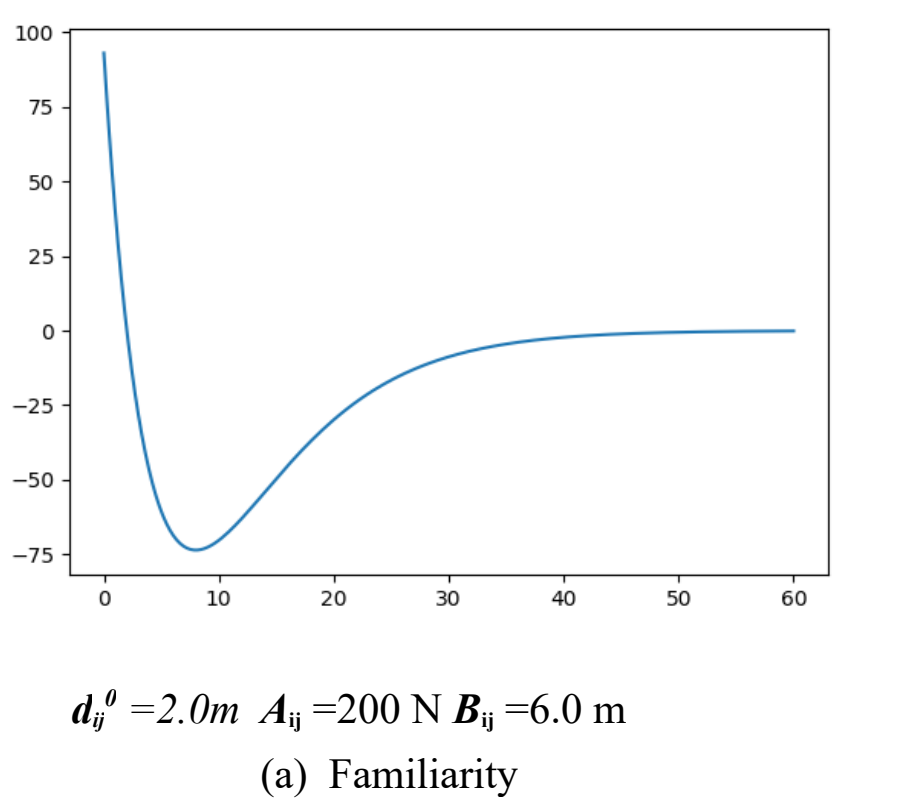

$d_{ij}^0$ =2.0m $A_{ij}$ =200 N $B_{ij}$ =6.0 m

(a) Familiarity

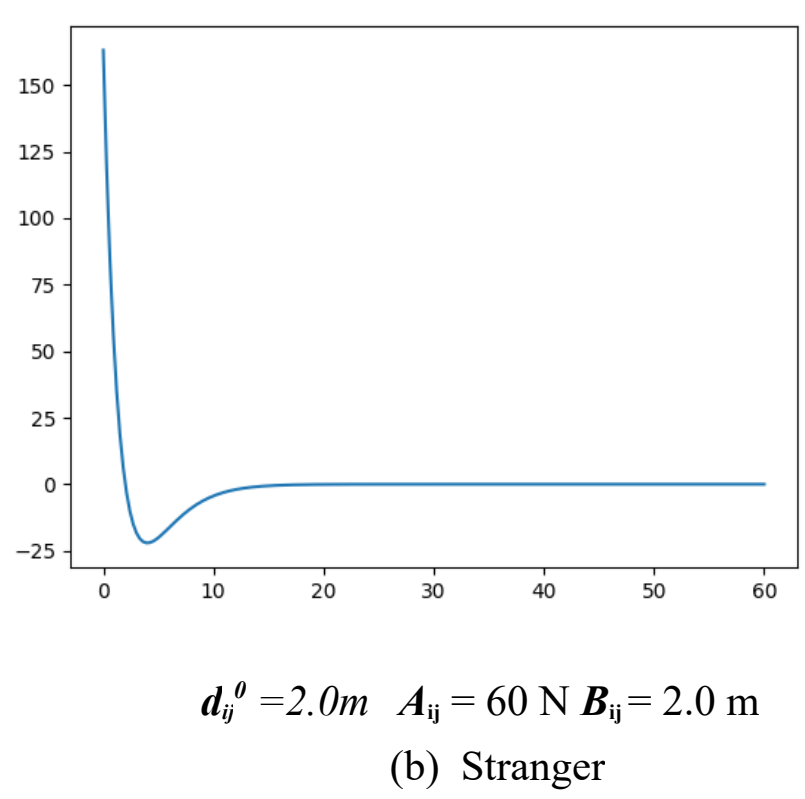

$d_{ij}^0$ =2.0m $A_{ij}$ = 60 N $B_{ij}$ = 2.0 m

(b) Stranger

**Figure 3.1. Extended social force from individual *j* to individual *i (non-anisotropic formula)*: (a) To characterize two individuals who know each other, the force includes a negative segment representing attraction as well as a positive segment representing repulsion; (b) When two individuals are strangers, attraction significantly decreases in both strength and the effective range.**

Moreover, the traditional formula of social force is compared with the group social force given by Equation (3.1), and it is noticed that the desired distance $d_{ij}^0$ is usually larger than $r_{ij}$, and parameter $A_{ij}$ and $B_{ij}$ are also in different values. In general, the traditional social force is usually considered as short-range interaction, and it plays a role of collision avoidance because it is calculated by using the physical size of individual agents (i.e., $r_{ij}$). In other words, the social force is effective only when people are very close to each other ($A_i$=2000; $B_i$=0.8), and it is fit to model high-density crowd. As for the new formula, it is relatively a long-range interaction where the desired distance $d_{ij}^0$ is commonly larger than $r_{ij}$, and parameter $B_{ij}$ of group social force is often larger than $B_i$ in the traditional formula. In our numerical testing, it is found that $B_{ij}$ is usually in the range of $10^1 \sim 10^{-1}$ while $A_{ij}$ is commonly in the range of $10^2 \sim 10^0$. This issue will be further discussed in detail in numerical testing results.

Equally importantly, the gap between $d_{ij}^0$ and $d_{ij}$ is expressed in Equation (3.1), and the interpersonal stress is characterized in consistency with our previous discussion. The gap of $d_{ij}^0$ and $d_{ij}$ is either negative or positive, meaning that being too far away or too close to someone result in stress in proximity. Keeping proper distance with others is the way to protect us from too much arousal, and this is evident in socio-psychological study because being isolated or overcrowded can both lead to stressful conditions.

In order to well apply the group social force in pedestrian modeling, there is another important issue referring to oscillation phenomenon when an individual pedestrian get close to the equilibrium position $d_{ij}^0$ when interacting with another one. A common method is to integrate the relative velocity $\boldsymbol{v}_{ji}$ =$\boldsymbol{v}_j$- $\boldsymbol{v}_i$ as a force component in Equation (3.1). This term is an estimate of the future trend of motion based on its current rate of change. It is sometimes called "anticipatory control," which characterizes a pedestrian is able to anticipate the changing situation for a short time, and thus adjust the current state of motion to avoid future collision or oscillation. This force component contributes to reducing oscillation when two agents get closely enough, and it has been applied in several other pedestrian models (Stenffen, 2008, Gao et. al., 2013). Thus, by tuning a force component that is a function of relative velocity $\boldsymbol{v}_{ji}$, the oscillation phenomenon will be significantly mitigated.

As above we briefly introduce $\boldsymbol{v}_{ji}$ =$\boldsymbol{v}_j$- $\boldsymbol{v}_i$ to offset oscillation. In practical computing even if the anticipatory force of $\boldsymbol{v}_{ji}$ is not used, we seldom observe oscillation. In fact the method of using relative velocity $\boldsymbol{v}_{ji}$ to adjust acceleration origins from vehicle traffic problems, where a vehicle adjusts the moving velocity based on the relative velocity to the front vehicle (Treiber, Hennecke, and Helbing, 2000). A major assumption is that vehicles are moving sequentially in lanes. However, human pedestrians are not often required to keep lanes in many occasions. In open areas such as in shopping malls or airports, pedestrians usually do not need to pay much attention to others' velocity unless they are very close to each other. Even if they are moving close, it is easy for a pedestrian to adjust moving direction in 2D space to avoid collision. This is a major difference between pedestrian movement in 2D space and vehicle movement in 1D space. Thus, although the force term of $\boldsymbol{v}_{ji}$ is often used in pedestrian simulation, it is not well justified theoretically to borrow this force term from vehicle traffic to pedestrian traffic.

$$d_{ij}^{0}(t+1)=(1-p_i)d_{ij}^{0}(t)+p_i d_{ji}^{0}(t) \qquad d_{ji}^{0}(t+1)=(1-p_j)d_{ji}^{0}(t)+p_j d_{ij}^{0}(t) \tag{3.2}$$

In real-world scenario people usually have a target interpersonal distance and get there without oscillation in physical positions. Thus, an important question is whether we should assume $\boldsymbol{v}^0$ and $d^0$ to be constant in the above analysis. As mentioned before, $\boldsymbol{v}^0$ and $d^0$ are not physical entities, but reflect pedestrians' opinions. If $\boldsymbol{v}^0$ and $d^0$ are not constant, pedestrians may adapt $\boldsymbol{v}^0$ and $d^0$ dynamically, and this feature may provide another method to avoid oscillation. Now we consider the group social force as given in Equation (3.1), and a typical example is that $d_{ij}^0$ and $d_{ji}^0$ interact such that two individuals become a stable pair in physical positions when interacting. In particular based on the opinion dynamics of Equation (2.1) we formulate the dynamics of $d_{ij}^0$ and $d_{ji}^0$ by Equation (3.2), and it is assumed that $c_{ij}$=1 when individual $i$ is only talking to individual $j$. Equation (3.2) can be viewed as a gossip-based asynchronous method for Equation (2.1). In Figure 3.2 we illustrate how $d_{ij}^0$ and $d_{ji}^0$ converge to the same value for two individuals.

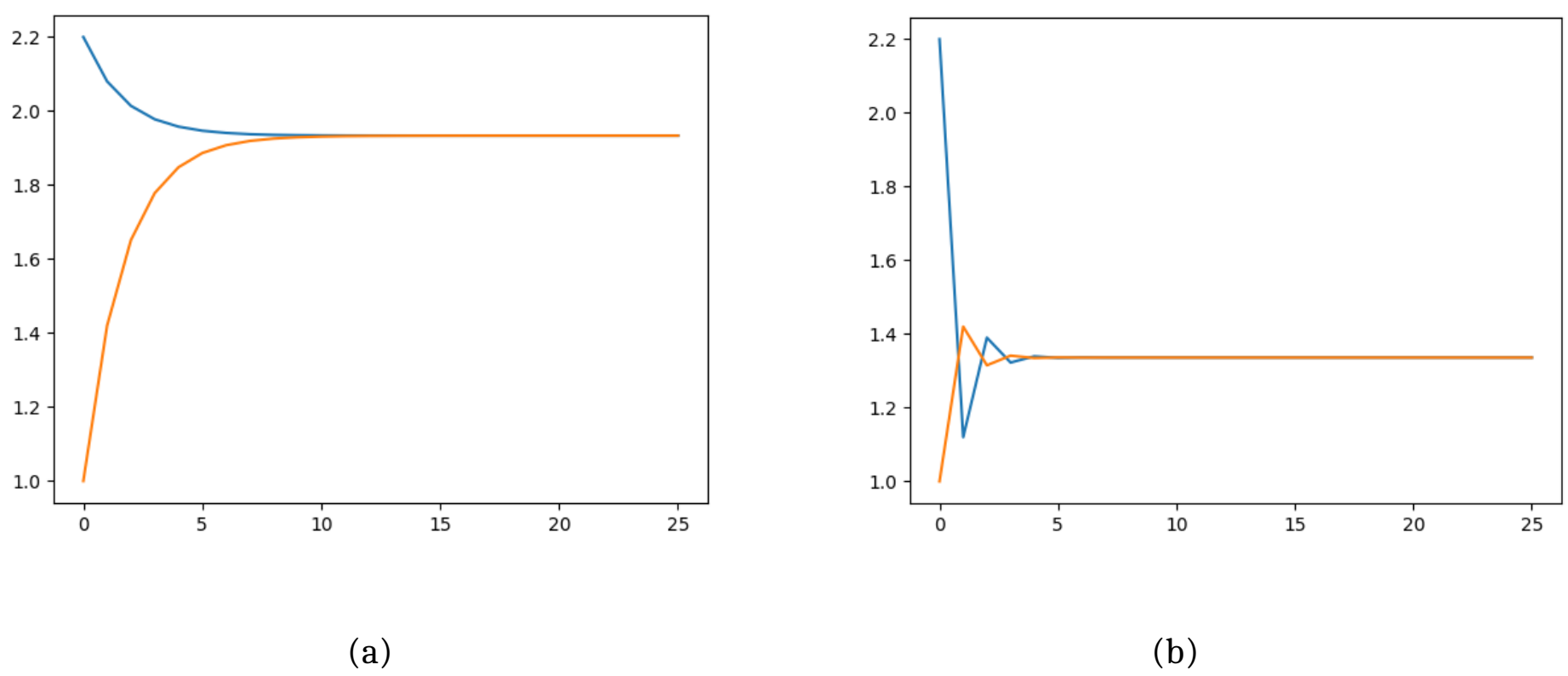

**Figure 3.2. Interaction of $d_{ij}^0$ and $d_{ji}^0$ by using opinion dynamics: $d_{ij}^0$ and $d_{ji}^0$ converge to a common value. The blue line indicates dynamics of individual *i* while the red line is for individual *j*. Initially $d_{ij}^0$ = 2.2 while $d_{ji}^0$ =1.0. In plot (a) $p_i$ = 0.1 and $p_j$=0.35 while in plot (b) $p_i$ = 0.9 and $p_j$=0.35.**

### (b) Social Group in Matrix-Based Structure

Based on the group social force by Equation (3.1), we are able to define a kind of matrix-based structure to describe social group behavior in crowd movement. Considering a group composed of $n$ individuals, the social relationship of the group members is described by a $n$x$n$ matrix S, of which the element is defined by $s_{ij}$ in Equation (2.5). In a similar manner we define $n$x$n$ matrices $D^0$, $A$ and $B$, of which the elements are $d_{ij}^0$ $A_{ij}$ and $B_{ij}$ as specified in Equation (3.1), respectively. Generally speaking, $D^0$, $A$ and $B$ are asymmetrical, and the matrix-based structure is given by Equation (3.3).

$$\boldsymbol{S}=[s_{ij}]_{n\times n} \qquad \boldsymbol{A}=[A_{ij}]_{n\times n} \qquad \boldsymbol{B}=[B_{ij}]_{n\times n} \qquad \boldsymbol{D^0}=[d_{ij}^{0}]_{n\times n} \tag{3.3}$$

The group social force among $n$ individuals is described by the matrices $D^0$, $A$ and $B$, and the method has been tested in our simulation platform CrowdEgress (Wang et. al., 2023) as well as FDS+Evac (Korhonen, 2018). To calculate the group social force exerted on individual $i$ by individual $j$, we pick up the element $d_{ij}^0$, $A_{ij}$ and $B_{ij}$ in the corresponding $n$x$n$ matrice $D^0$, $A$ and $B,$ and apply mathematical formula of Equation (3.1) to compute $\boldsymbol{f}_{ij}^{soc}$. Here the social structure of $n$ individuals is also characterized by matrices $D^0$, $A$ and $B$. Because both the opinion dynamics and group social force describe social relationship and such two features are inter-related, and it is important to define A, B, $D^0$ in consistency with S.

A testing result of FDS+Evac is illustrated in Figure 3.3, where the multi-compartment layout is given based on an example in Pan et. al., 2007. Several groups are well observed in the simulation result. Some small groups merge into a large group and regrouping occurs at intersections or bottlenecks. In general, the group concept in the matrix-based structure is not static, and we do not need to define the group members in a static approach. Rather, we only define the social topology of $n$ individuals, and their interaction forces will determine how such $n$ individuals are self-organized into groups. Furthermore, the elements in matrices $D^0$, $A$ and $B$ could be time-varying as exemplified by Equation (3.1), and the group could be restructured dynamically through the simulation process, and thus it inherently becomes a time-varying feature.

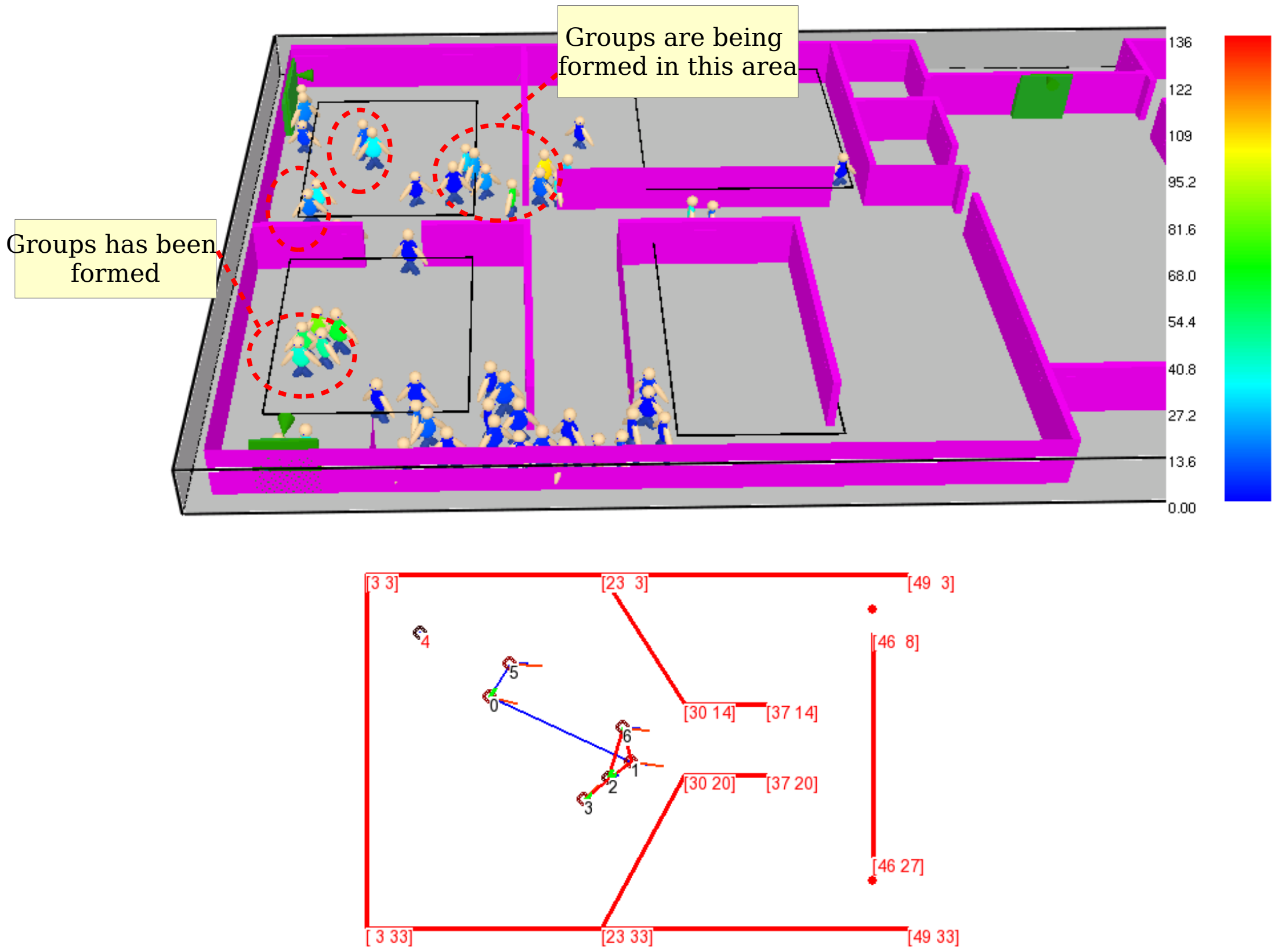

**Figure 3.3. Simulation of Group Dynamics in FDS+Evac: The color bar indicates the intensity of the group social force given by Equation (3.1). An individual in active motion usually moves in the front of a group. Individuals in passive motion are followers in the group and they usually move behind the leaders.**

Based on Equation (3.1) and (3.3), a typical pattern of crowd movement is described as the leader-and-follower group. In this group pattern there is a kind of individuals whose behavior is mainly motivated by himself or herself. If others would like to follow them, they become leaders in group behavior. Thus, if individual $i$ is the leader in a group, his or her motion is primarily motivated by self-driving force. In contrast followers are those whose behavior is mainly motivated by others, and their motions are thus mainly governed by group social force, and the self-driving force is secondary. As mentioned before imbalance (asymmetry) of $d_{ij}^0$ and $d_{ji}^0$ will contribute to model leadership in crowd behavior. In a similar manner, if $i$ is a leader, $B_{ji}$ is usually larger than $B_{ij}^0$ such that the leader will attract his surrounding people, but not easily be attracted by them. In brief an individual's motion can be classified into two types. One type of motion is primarily motivated by the self-driving force, and is called active motion. The other type of motion is largely motivated by surrounding other individuals, which is called passive motion. Usually an individual's motion is a combination of both types, but we can differentiate such two types in simulation results, and identify whether an individual's motion is dominated by either active or passive type.

As shown in Figure 3.3 the color bar is used to observe the magnitude of the group social force in simulation (Korhonen 2017; Forney, 2018; McGrattan et al., 2018). An individual in active motion often moves in the front of a group. Individuals in passive motion are followers in the group and they usually move behind the leader. The leader is commonly under smaller group social force than the followers.

### (c) Social Groups and Opinion Dynamics

Social groups and opinion dynamics are related but different concepts. The group social force describes the social relationship of individuals and combine them together in physical positions. When people are sufficiently close to each other, herding effect and opinion dynamics becomes effective such that their opinion interact to form a common motive or move towards a common destination. This is due to the criterion of $d_{ij}<R_i$ as introduced above, namely individual $i$ will be able to know surrounding people's opinions within the range of $R_i$ by talking or observing their behavior (See Equation (2.1)). Furthermore, one's opinion is more affected by those who have close social relationship, such as family members or friends, and thus the interaction range is not only determined by physical distance $d_{ij}$, but may also refer to the social distance $d_{ij}^0$, and social typology of individuals becomes an important factor. For example we may assume that opinion of individual *i* is impacted by individual *j* if $d_{ij} < d_{ij}^0 + 2B_i$ .

As social groups are jointly modeled with opinion dynamics, a kind of convergent pattern is supposed to emerge in a crowd. A common example is evacuation of a stadium where people follow the crowd flow to move to an exit. There are a multitude of small groups composed of friends or family members, and they keep together in egress because of their social relationship. These small groups compose a large group of evacuees, and herding behavior widely exists among these groups, contributing to form a common motive of motion. In sum, the group social force makes individuals socially bonded with each other, and it primarily characterizes a kind of social relationship of individuals. Herding effect emphasizes how much an individual tends to follow others' opinion or behavior, and the effect is also related to their social relationship, and it is common to assume that an individual is more inclined to follow those who are in closer relationship. Because both social groups and herding effect considers social relationship and such two features are inter-related, and it is important to define them in consistency. In other words, the matrices $D^0$, $A$ and $B$ should be defined in consistency with matrices $P$ and $C$. In practical computing, it is important to first define matrix $C$ and $P$ as a quantitative measure of social relationship, and then specify $D^0$, $A$ and $B$ in consistency.

Moreover, the group social force as presented above is also useful to model crowd behavior in pre-movement phase in crowd evacuation (Sorensen, 1991, Kuligowski, 2009). In brief, when the alarm or hazard is detected, people usually do not head to exits immediately, but go to find trust ones (e.g., family members or friends) for information exchange and collective decision. Such social group behavior delays their movement to exits (Lovreglio and Kuligowski, 2022). Thus, the above model of social groups contributes to modeling the crowd behavior in pre-movement phase and will be useful to investigate how the initial delay is formed and influenced by the group dynamics. In particular Equation (3.1) can be better explained by flight-or-affiliation effect in pre-movement phase of evacuation. The self-driving force motivates one to flee while the group social force makes one affiliated with others. This effect agrees with social attachment theory in psychological study (Mawson, 2007; Bañgate et al., 2017). The social attachment theory suggests that people usually seek for familiar ones (e.g., friends or parents) to relieve stress in an emergency situation, and this is rooted from our instinctive response to danger in childhood when a child seek for the parents for shelter. However, such social attachment could take some time and it might delay the movement towards exits. Thus, different from the fight-or-flight response (Cannon, 1932), the group social force agrees with the flight-or-affiliation effect, especially in the pre-movement phase.

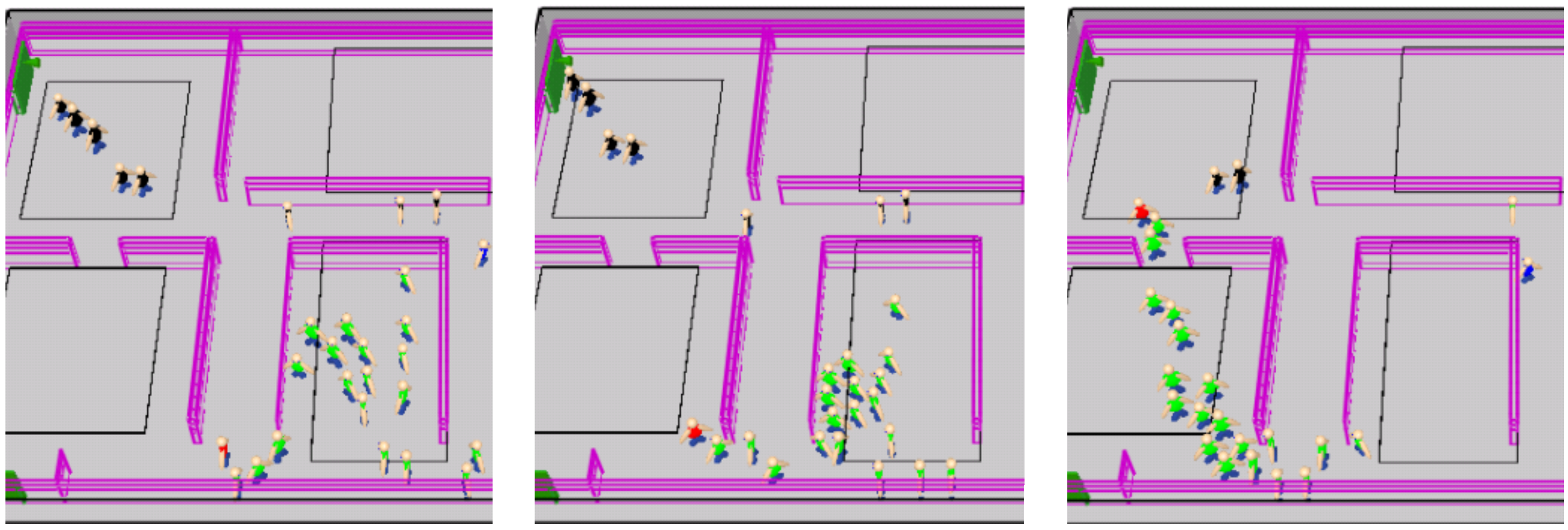

**Figure 3.4. Herding Effect in Pre-Movement Phase and Movement Phase in a Multi-Compartment Layout. The structural layout is created from MassEgress project (Pan, 2006), and the numerical testing case is realized by using NIST simulator of FDS+Evac (Version 6.5.3) and Smokeview. The source code of FDS+Evac is modified to include the above pre-evacuation model and group social force.**

As above we illustrate a simulation result of FDS+Evac where herding effect is jointly used with social groups in the pre-movement phase. The group social force is applied as introduced in our previous work (Wang, 2016) so that agents are attracted to each other if they are in close relationship, and their opinions interact when they are physically close enough, namely $d_{ij} < R_i$ . Here the red agent has *tpre* earlier than the green ones, and Equation (2.5) is effective to reduce the green agents' *tpre*. As a result, the pre-movement time converges for agents in both colors. Moreover, the group social force combine them together in physical positions. When the simulation time goes beyond their common *tpre*, the entire group of green agents tends to move with the red one towards an exit, and the leader-and-follower pattern appears due to the imbalance (asymmetry) of $d_{ij}^0 \neq d_{ji}^0$ between agents in two colors.

In the following discussion the criterion is extended by assuming that opinion of agent *i* is impacted by agent *j* if $d_{ij} < d_{ij}^0 + 2B_{ij}$ . In other words, agent *i* merges opinion with agent *j* when their physical distance are within the effective range of group social force. This extended criterion implies that opinion dynamics and group social force take effect together to form social group behavior, and it is being tested in our simulation platform CrowdEgress. In CrowdEgress $d_{ij} < d_{ij}^0 + 2B_{ij}$ is used to generate an attention list for each agent, and this list is used to update the value of $b_{ij}$ . The connectivity of a time-varying social topology is described by $[b_{ij}]_{nxn}$.

The social relationship of 8 agents are specified in Table 3.1, where the matrices of $\boldsymbol{S}$, $\boldsymbol{A}$, $\boldsymbol{B}$, $\boldsymbol{D}^0$ are sequentially defined. The agents are sequentially identified by their index number i=0,1,2...7. The major group is formed by agent 0, 1 and 2, which is similar to the previous pre-movement example as introduced in Section 2. In this group agent 0 is completely a follower to agent 1 with a social weight of $s_{01}$=1.0, and agent 1 cares about agent 0 with a social weight of $s_{10}$=0.2. Agent 1 and 2 are socially connected with each other with $s_{12}$=0.5 and $s_{21}$=0.7, and both of them are also socially connected with agent 3 with weight $s_{13}$=0.3 and $s_{23}$=0.3.

**Table 3.1. List of Group Parameter S|A|B|D.**

<table>
<tr><th>&groupSABD</th><th>Agent0</th><th>Agent1</th><th>Agent2</th><th>Agent3</th><th>Agent4</th><th>Agent5</th><th>Agent6</th><th>Agent7</th></tr>
<tr><td>Agent0</td><td>0</td><td>1|170|10|2</td><td>0</td><td>0</td><td>0</td><td>0</td><td>0</td><td>0</td></tr>
<tr><td>Agent1</td><td>0.2|60|6|2</td><td>0</td><td>0.5|30|10|3</td><td>0.3|20|20|1</td><td>0</td><td>0</td><td>0</td><td>0</td></tr>
<tr><td>Agent2</td><td>0</td><td>0.7|30|20|3</td><td>0</td><td>0.3|90|3|2</td><td>0</td><td>0</td><td>0</td><td>0</td></tr>
<tr><td>Agent3</td><td>0</td><td>0</td><td>0.3|30|6|2</td><td>0</td><td>0</td><td>0.1|10|2|1</td><td>0.6|60|2|1</td><td>0</td></tr>
<tr><td>Agent4</td><td>0</td><td>0</td><td>0</td><td>0.2|70|20|1</td><td>0</td><td>0.3|20|2|1</td><td>0</td><td>0.5|20|3|1</td></tr>
<tr><td>Agent5</td><td>0</td><td>0</td><td>0</td><td>0</td><td>0</td><td>0</td><td>1|70|3|1</td><td>0</td></tr>
<tr><td>Agent6</td><td>0</td><td>0</td><td>0</td><td>0</td><td>0</td><td>1|200|1|1</td><td>0</td><td>0</td></tr>
<tr><td>Agent7</td><td>0</td><td>0</td><td>0</td><td>0.5|26|8|1</td><td>0</td><td>0.5|3|1|1</td><td>0</td><td>0</td></tr>
</table>

In contrast, agent 5, 6 and 7 form another social group. In specific agent 5 and 6 compose a stable pair because they are influential in mutual and exclusive manner. Agent 4 and agent 7 also follow agent 3 and 5, but agent 5 does not care them, and thus such a social relationship is not balanced.

Agent 3 and 4 connect both groups, which is similar to the previous example in Section 2. Such connection are directional in a sense that agent 3 and 4 are impacted by agent 5, 6, 7 (second group) and agent 0, 1, 2 (first group) are moderately impacted by them. Thus, in the above social topology agent 5 and 6 seems to maintain a kind of leadership, who directly impact agent 3, 4, 6 and 7 and relies on agent 3 and 4 to indirectly affect agent 0, 1, 2. In this social topology agent 3 critically connects the two groups. Without agent 3 the two groups become isolated completely.

Given the above social relationship weights, the above eight agents are simulated in a multi-compartment layout to determine their pre-movement time in Figure 3.5. The compartment layout is created based on an example in Wang et. al., 2008. The nearest-exit strategy is used in exit selection algorithm, and the pre-evacuation time are simulated by using the above matrix-based social structure. There are three other factors of importance in determining the dynamics of pre-movement time:

(1) Parameter $p_i$ is critical to determine how an individual keeps balance between his or her own opinion and others' opinions in the given social topology. The value of $p_i$ for the above agents is given by 0.6, 0.6, 0.3, 0.3, 0.4, 0.36, 0.63 and 0.66. As mentioned before, a small $p_i$ value implies that an agent is stubborn such that he or she would not like to accept opinions from others. In contract a large $p_i$ value means that an agent has an open mind and tends to hear or follow others' opinions. This feature probably refers to one's personality, and it is feasible to infer $p_i$ from existing psychological research.

(2) Initial positions of agents are illustrated in Figure 3.5. The initial positions significantly affects how agents move and interact with others in the simulation process. Agents in different compartments cannot have social interaction no matter how $s_{ij}$ is specified. As mentioned previously, opinion exchange as described by Equation (2.1) and (2.5) only becomes effective within the range of $R_i$ , namely, $d_{ij} < R_i$ . In the following study case we specify $R_i$ by $d_{ij}^0 + 2B_{ij}$ , and opinions exchange when $d_{ij} < d_{ij}^0 + 2B_{ij}$ such that opinion dynamics and group social force take effect together to form social group behavior. Thus, the physical positions of agents matters in the context of social interaction.

(3) Initial pre-movement time $tpre_i$ is given by 3, 2, 10, 6, 22, 6, 16 and 23 (minutes) sequentially. In existing research it is often assumed that pre-movement time follows certain probability distributions, such as Lognormal, Loglogistic or Weibull distribution (Forssberg et. al., 2019). Such statistical study is useful to explore whether or to what extent the pre-movement time is related to other factors such as the crowd size, existence of a leader, or the location of evacuation (e.g., railway stations or libraries). Thus, reliable data may provide a guideline to initialize the pre-movement time in simulation. However, a difficulty exists in acquiring reliable data in real-world events, and how to initialize $tpre_i$ may also refer to psychological studies on this research topic.

In Figure 3.5(b) each colorful line represents $tpre_i(t)$ for each evacuee and the they converge into two groups as evacuees interact with each other. When the colorful line drops to $t$ axis, it means that the evacuee reaches a safe place and thus is removed from computational loop. The figure is plotted by using our simulation platform CrowdEgress. As mentioned before, a dash line divides the plane equally into two regions. The upper left region is the pre-movement phase of $tpre_i>t$, and the lower right region is the movement phase for $t \geq tpre_i$. When the colorful lines meet the dash line, state transition happens from pre-movement phase to the movement phase.

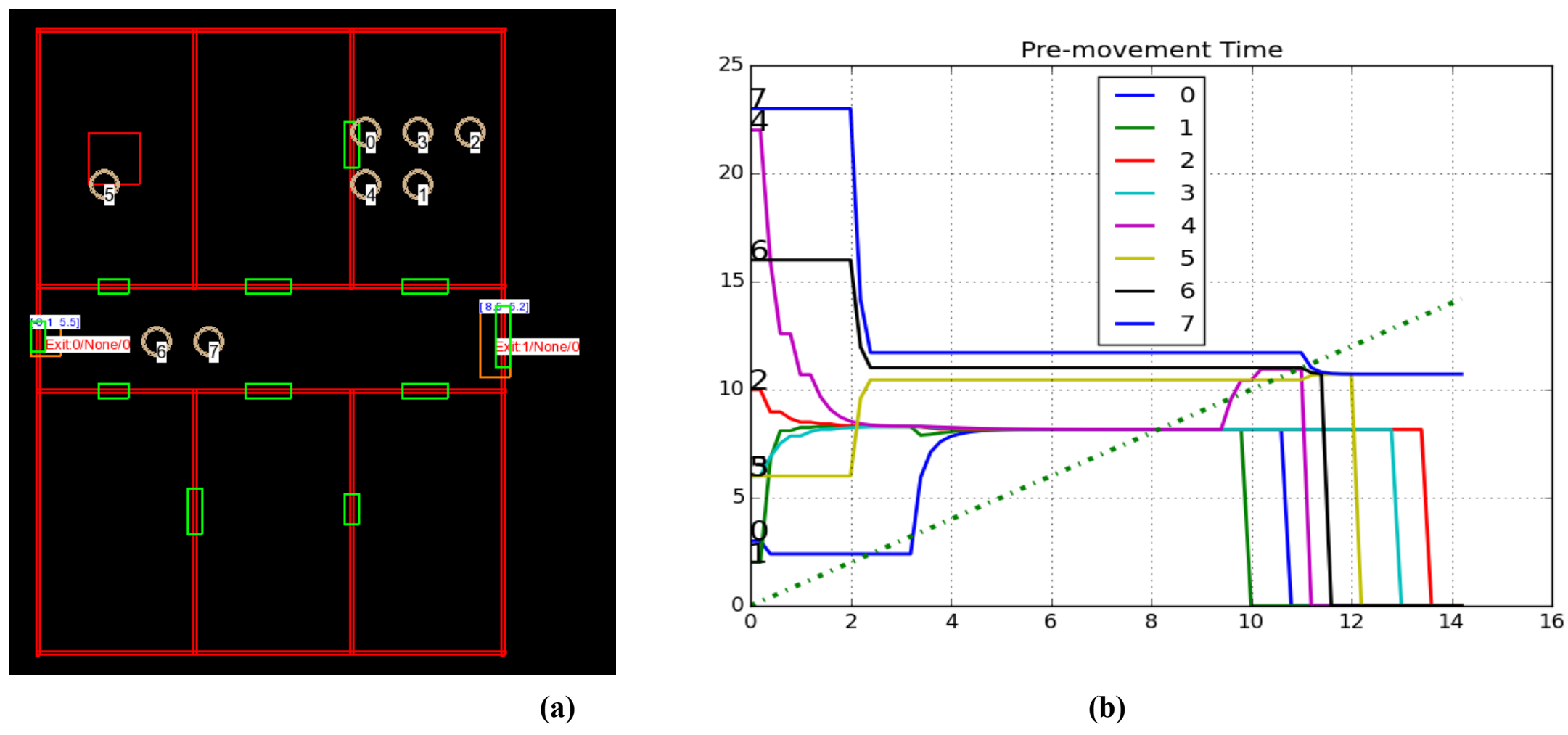

(a) (b)

**Figure 3.5. Opinion Dynamics of Pre-Movement Time: The example is created from our IEEE Conference paper (Wang et. al., 2008), and the numerical testing results including input and output data can be downloaded at https://sourceforge.net/p/crowdegress (in the folder of /examples/TASE_Conferece2008).**

As mentioned before in Section 2, we are more interested in the upper left region, where we are able to identify how social interaction occurs among these eight agents. The convergence of pre-movement time is illustrated in Figure 3.5(b). Because agent 5 and 6 exclusively care about each other and do not follow any others, they compose a group and their pre-movement time converge to roughly 11 min as shown in Figure 3.5(b). In contrast $tpre_i$ of agent 0 to 3 quickly converge to 8 min. The curves indicate that agent 4 communicate with agent 5, 6 and 7 after 9min, and its pre-movement time converges towards the second group around 11min.

In a sense the matrix $W=[w_{ij}]_{nxn}$ induced from $c_{ij}$, $b_{ij}$ and $s_{ij}$ provides a whole picture to understand how individuals make decisions. In a mathematical sense $W=[w_{ij}]_{nxn}$ defines a time-varying social structure to interpret how individuals' opinions evolve in social interactions, and such a system dynamics is formulated by $OPIN(t+1)=W \cdot OPIN(t)$. Matrix $W=[w_{ij}]_{nxn}$ is updated as individuals move and interact. Thus, in above example individual 3 and 4 can also join the second group with agent 5, 6 and 7. Such simulation results are both possible and reasonable, and it depends on the initial positions of these agents.

## 4. Way-Finding Activities

Way finding refers to how people orientate themselves towards exits within a multi-compartment building, and it involves perception and integration of various information acquired from surrounding facilities and individuals (Tong and Bode, 2022). The problem could be formalized in two steps: (a) How to choose a destination exit; (b) How to select a proper route to reach the exit selected. There are two major stressors that influence their decision making in way-finding activities. A major stressor is the guidance information received such as exit signs or voice information acquired from surroundings, and they usually helps to reduce the safe egress time. Another stressor is perception of hazard threat as discussed in the previous section. For example, if evacuees perceive smoke or heat on their ways to an exit, they probably will consider adjusting their escape route or simply move to another exit.

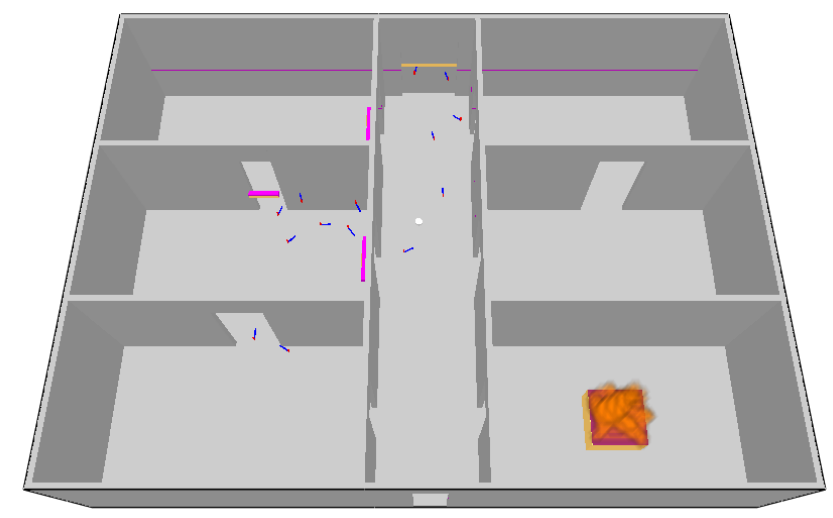

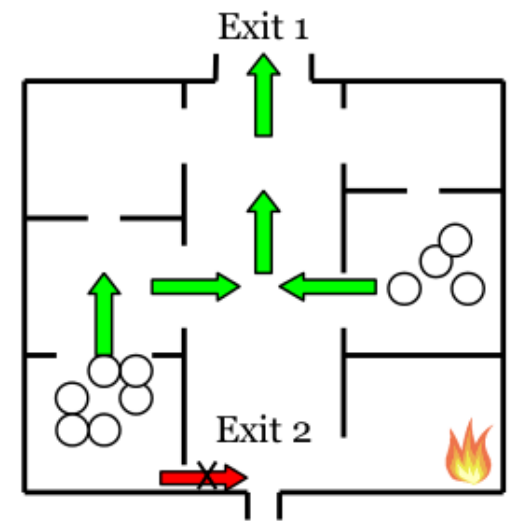

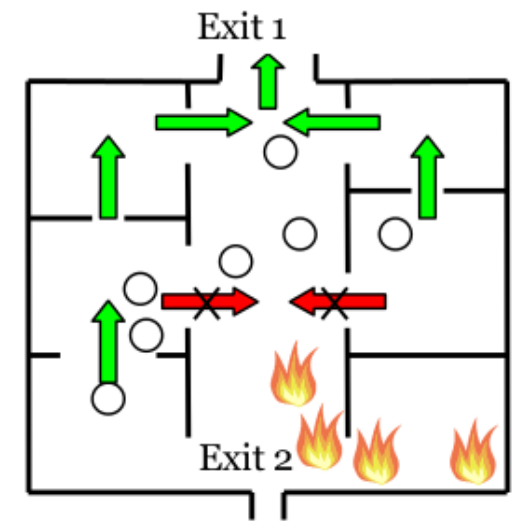

**Figure 4.1. Way-Finding Activities and Exit-Selection Behavior: In this case the exit and route are dynamically updated in the time line of evacuation process, and the route is recalculated when the fire and smoke spread into the corridor area. The example is illustrated from our IEEE Conference paper (Wang et. al., 2008)**

Mathematically, it is reasonable to use a probabilistic model to describe such way-finding activities by integrating all the information available for evacuees (Wang et. al., 2008, Zheng et. al., 2015). Take the above figure for example, and there are two choices, and each evacuee is assigned with probability $[\mu_1, \mu_2]$ to select either exit 1 or exit 2. The prior probability distribution $[\mu_1, \mu_2]$ is given as the normal usage of the exits, and it can be given as [0.5, 0.5] if there is no bias preset in people's mind, assuming the two exits are equally used in normal life. In principle the prior indicates the historical usage frequency of different ways on daily basis and it reflects a kind of way-selection habit of people. Such frequency-based data provides a reliable estimate of $[\mu_1, \mu_2]$ based on the law of large numbers in probability theory, and it well carries the information that people are inclined to use the familiar exits rather than unfamiliar ones (Proulx, 1993). The effect of fire drills or other training events may also be integrated in this prior probability distribution. For instance, evacuees ever participated in a fire drill where Exit 2 was used as fire escape route, and such past experience will dramatically increase $\mu_2$ in initial estimate of the probability distribution.

In the simulation process the prior probability distribution of using multiple exits will be dynamically updated when new information received. In Figure 4.1, when the evacuees observe that the fire and smoke spread into the corridor, it will be reasonable to update the probability distribution, for example, to be [0.9, 0.1] such that exit 1 is mainly used for safe egress. Consequently, it is feasible to build up a probabilistic graph model to link different factors in a statistical sense (Wang et. al., 2008). In addition, if evacuees find an exit sign at any intersections, they will balance it with their own knowledge or habit, and choose to follow the sign or not. This process refers to how an individual deals with the new information received from surroundings and make decisions. In existing social-economic systems utility theory is very useful to integrate various factors into one's decision making process (Tong and Bode, 2022), and it provides a quantitative measure on how useful a choice is to an individual. In the above exit-selection problem the utility of exit $q$ to individual $i$ consists of a deterministic component $V_{iq}$ and a random component $\varepsilon_{iq}$. The deterministic component, for example, could be expressed as below.

$$V_{iq} = \pi_1\,\mathrm{DIST} + \pi_2\,\mathrm{SM} + \pi_3\,\mathrm{EL} \tag{4.1}$$

In Equation (4.1) DIST is the distance of the individual agent $i$ to exit $q$ (e.g., Eularian Distance or Manhatten Distance). SM is a boolean variable indicating if there is any smoke or hazardous materials perceived by agent $i$ on the path to exit $q$, and EL is another boolean variable indicating if there is any evacuation lights or exit signs detected by agent $i$ on the path to exit $q$. The sub-level parameters in Equation (4.1) include $\pi_1$, $\pi_2$ and $\pi_3$. To calibrate the sub-level parameters, random utility theory (RUT) is used based on data collected from real-world experiments, and according to Lovreglio, Fonzone and dell'Olio, 2016 and Edrisi, Lahoorpoor and Lovreglio, 2021 the above sub-level parameter are given by $\pi_1$= -0.011, $\pi_2$=-0.985 and $\pi_3$=0.175. Combining prior information with a utility function we formalize a probability function of discrete choices as below.

$$\text{Prob(individual } i \text{ select exit q)} = a_1\, prior_{iq} + a_2 \frac{\exp(V_{iq})\,kw_{iq}}{\sum_k \exp(V_{ik})\,kw_{ik}} + a_3\, others(q) \qquad \text{and } a_1+a_2+a_3=1 \tag{4.2}$$

where $V_{iq}$ is the utility function of exit $q$ for individual $i$. Here $kw_{iq}$=1 if individual $i$ knows exit $q$, and $kw_{iq}$=0 if not. It is also necessary to ensure the model to satisfy that $prior_{iq}$= 0 if $kw_{iq}$=0, i.e., $prior_{iq}$=0 if individual $i$ does not know exit $q$. In other words, there is a consistency problem in updating the related terms in the right side of Equation (4.2). For example, $kw_{iq}$ should be consistent with $prior_{iq}$ such that $prior_{iq}$ is non-zero only if $kw_{iq}$ =1. If others tell individual $i$ that there is exit q, then $kw_{iq}$ becomes non-zero, but $prior_{iq}$ remains zero because individual i never go there before, and thus has no past experience of using the exit. In brief the above utility model reflects the idea of bounded rationality, which comes to dominate in most theories of individual decision making, especially in social-economic process. Such a decision process is "bounded," because usually not all options are known to the decision-maker.

Next, we will expand the above model in timeline by introducing the memory effect. As a result, $prior_{iq}$, others(q), $V_{iq}$ and $kw_{iq}$ are all functions of time $t$, which are timely updated in a consistent manner through the simulation process. In specific $prior_{iq}(t)$ represents the preference of individual $i$ to follow past memory on selecting the alternatives $q$, and the memory effect is formalized by $prior_{iq}(t)= \sum_z \text{Prob}_{iq}(z)$ with $z$=0, 1, … $t$-1. Here $\text{Prob}_{iq}(t)$ denote the probability of individual $i$ to select exit $q$ at time $t$ and $prior_{iq}(t)$ becomes a linear combination of $\text{Prob}_{iq}(z)$ from $z$=0 to z=t-1. A most simple method is using Markovian property where $prior_{iq}$ is iterated by $prior_{iq}(t)=\text{Prob}_{iq}(t-1)$.

More importantly, Equation (4.2) is well consistent with the opinion dynamic model as introduced in Section 2, where $a_3=p_i$ and it indicates how much an individual tends to follow others' characteristics. The $opinion_i(t)$ in Equation (4) is replaced by a probability distribution in Equation (4.2), which represents an individual's selective preference of using different exits, and it is timely updated as the individual interacts with others or receives new information from egress facilities. Based on Equation (4.2) a collective decision-making problem is formulated, where exchange of individual opinions could lead to either one large group (consensus) or several small groups (clustering).

Based on the probability chain rule and Equation (4.2), each individual's decision is made by integrating three factors: (1) historical knowledge of a structural layout or prior information such as habit of using certain path; (2) collecting timely information and judgment on current situation; (3) learning from opinions of surrounding others. The above three factors are balanced by tuning parameter $a_1$, $a_2$, $a_3$.

In a mathematical sense parameter $a_1$, $a_2$, $a_3$ are real numbers normalized in range of [0, 1], indicating one's preference of either using prior experience or timely information for decisions, or simply following others' decisions. They could also be understood as probability measure based on the total probability formula. In a psychological sense the parameter $a_1$, $a_2$, $a_3$ critically refers to one's personality, such as whether an individual is stubborn or open-minded. We will elaborate this issue in detail as follows.

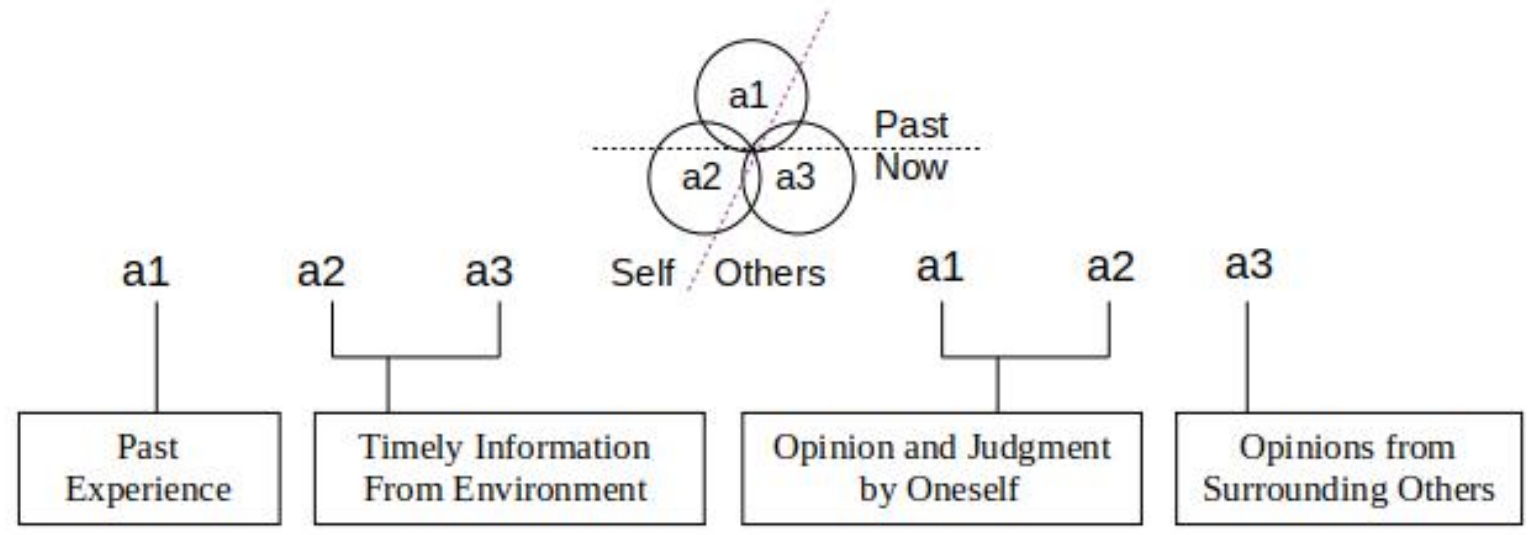

**Figure 4.2. Meaning of parameter $a_1$, $a_2$, $a_3$. The parameter $a_1$, $a_2$, $a_3$ indicates the confidence level of one's past experience or knowledge, one's own judgment on current situation as well as learning from opinions or decisions of others. Here a1, a2 and a3 are individualistic features, meaning that their values could be assigned differently for each individuals. In a mathematical sense a1, a2, a3 are real numbers normalized in range of [0, 1], indicating one's preference of either using prior experience or timely information for decisions, or simply following others' opinions or decisions.**

Increase of $a_1$ means that an individual relies heavily on his/her past experiences to make decisions while current situation is not specifically investigated. Because many life events are highly repetitive, it is common for an individual to compare current situation with their past experience and follow a set of routines to deal with the new situation. This behavioral pattern is straightforward and quick, and it does save time. In emergency egress, it is widely recognized that an individual's experiences can significantly impact his/her behavior, such as the familiarity of the surroundings, safety procedures, and fire drills. For example, evacuees are normally inclined to use their familiar path in egress, and such familiarity is integrated in the prior information as historical frequency of using different exits in normal situations, and parameter $a_1$ could be understood as one's probability of using such familiar path in egress. In a psychological sense $a_1$ increases if one is stubborn to their daily routine and not open-minded to new choices.

However, using prior evacuation experience to guide evacuation decisions, may or may not produce better outcomes, and a typical example is that most people tend to follow the route that they are most familiar with, and ignore alternate routes. The possible result is that they create a bottleneck path collectively as mentioned in Section 2. Thus, we also need parameter $a_2$ to describe a kind of bounded rationality by investigating timely situation. Increase of $a_2$ implies that an individual relies relatively less on their past experience, but tend to investigate current situation to make a decision. For example, an evacuee tends to use the nearest exit rather than normally-used path to reach the safety, and this decision process requires comparing the distance from the individual to all known candidate exits, and selecting of the optimal solution. The rational decision is widely studied in many social-economic systems, and it is mathematically described by discrete choice process with utility function $V_{iq}$. In our modeling framework parameter $a_2$ indicates how much one prefers to use such rational choice. Such rationality is "bounded" because usually not all options are known, and thus we have parameter $kw_{iq}$ to describe if option $q$ is known to individual $i$.

The parameter $a_3$ is of significant importance because it characterize how people will interact with each other to reach a collective decision instead of an individual decision. This is especially useful to form a social group in crowd as introduced in Section 2, and also it is evident in most of egress events that people usually tend to stay together to gain a sense of safety. Thus, to reach a collective decision is more widely observed and meaningful than individual decision making, and people are thus self-organized into social groups to respond to the outside environment stressors.

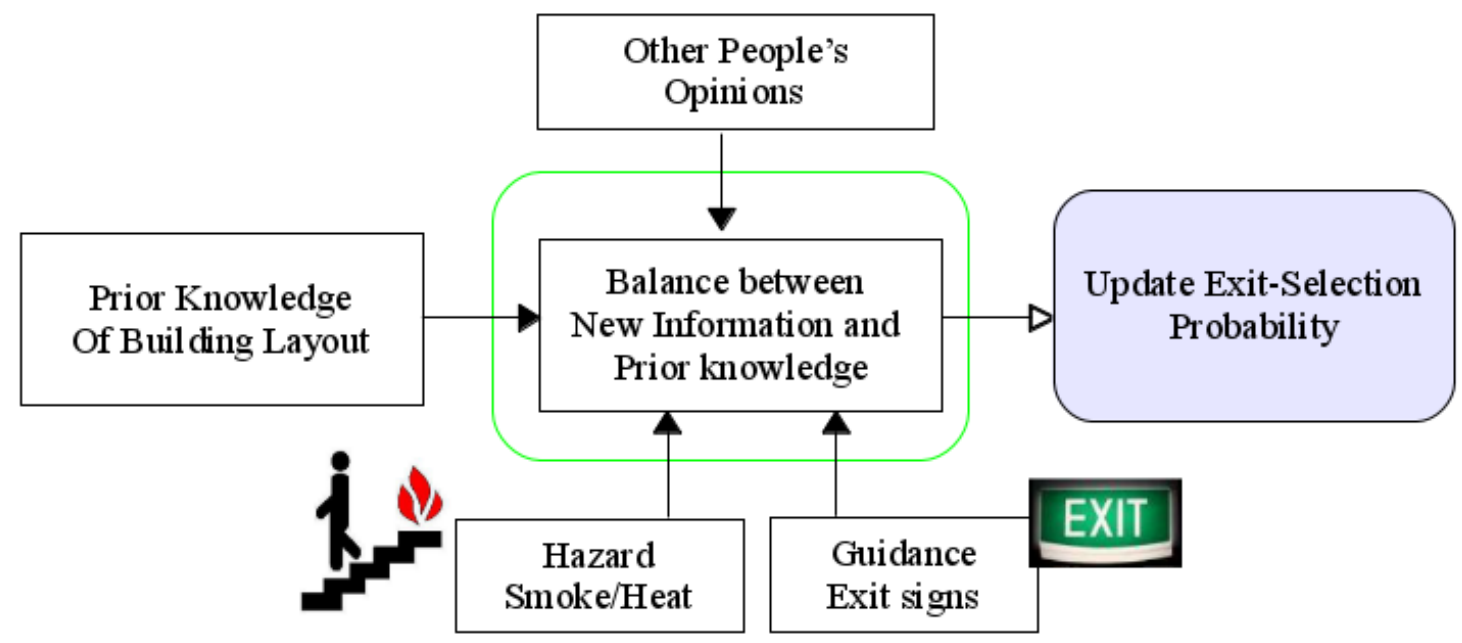

**Figure 4.3. Probabilistic Graphic Model of Way-Finding Activities: The exit-selection probability integrates various information including people's prior knowledge on building structure and location of exits. The probability distribution is updated if new guidance received or smoke detected.**

The value of $a_1$, $a_2$, $a_3$ could be formalized by using Analytic Hierarchy Process (AHP), which is a method for organizing and analyzing complex decisions, using math and psychology (See Appendix for more details). In sum the probabilistic model is useful to describe psychological findings on people's way-finding behavior, and a graphic structure is shown in Figure 4.3 to describe people's way-finding behavior. In fact analytic hierarchy process (AHP) and utility theory were initially learned from economic models. Different from the original use of such models in social-economic field, these models are partly renewed to capture various social-psychological findings in evacuation study. This study topics refers to an interdisciplinary field called social-physics or economic-physics (Quang et. al., 2018).

The above model is developed based on the assumption that crowd are mainly composed of rational individuals who require a sense of safety. Such a sense of safety is obtained by either simply following others or confidence on one's own judgment. In fact a small amount of panic individuals, or as called irrational behavior, could be also added as noise in the right side of Equation (4.2), but we will not further discuss this issue in this article. In brief Equation (4.2) enables us to formulate a collective decision making process, and the problem is interesting because we discuss how individual decisions interact with each other and whether or in what condition they converge to form the group opinion.

Now take a typical two-exit structural layout for example. It is shown in Figure 4.4 that exit selection probability could be either converged or polarized as many individual interact in given social topology. Take agent 2 for example. The probability for agent 2 to select the exit in the figure is initialized with 0.5. Through the simulation process the probability converges towards others and composes a major group of 1, 2, 3. In practical computing the way-finding activities mainly affect desired velocity $\boldsymbol{v}_i^0$ in our model. There are normally two steps involved: exit selection and route calculation.

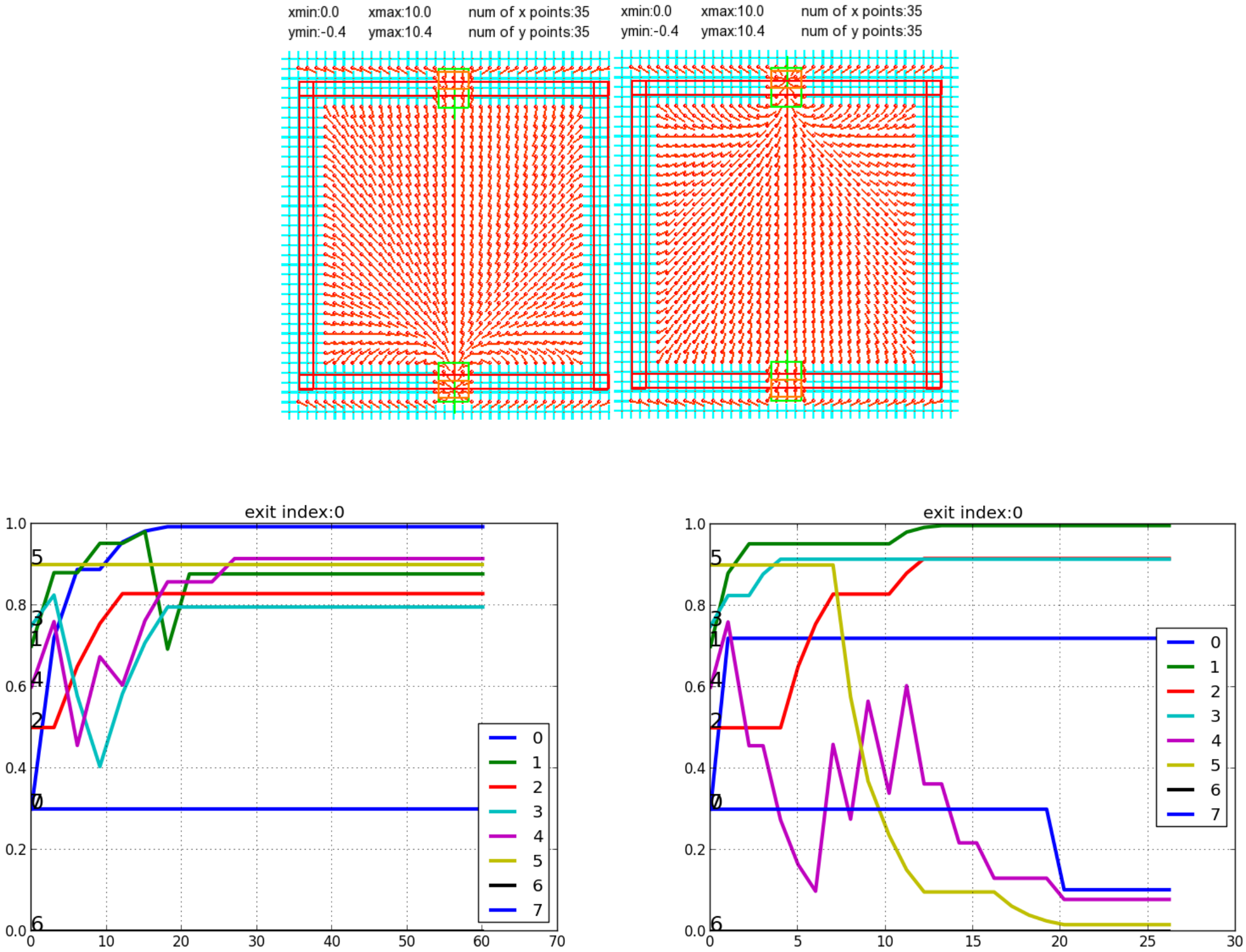

**Figure 4.4. Simulation of crowd egress with exit-selection algorithm: A binary decision problem is mainly investigated for exit selection algorithm, and probabilistic model are applied to describe how evacuees select an exit from two alternatives. The numerical testing results including input and output data can be downloaded at https://sourceforge.net/p/crowdegress (in the folder of /examples/OpinionModel_2Choices).**

The first step is to select an exit known by evacuees. This step is based on a probability model as mentioned above, integrating various information available for evacuees, including exit-selection habit as well as the guidance received and hazard perceived in emergency egress. In CrowdEgress and FDS+Evac all the exits are classified in two types: known exits and unknown exits. It is natural to assign each unknown exit with zero probability, which means that people never select an exit that they do not know. However, a known exit can also be assigned with zero probability. As mentioned before, if we consider the probability $prior_{iq}$ for q=1, 2, … Q indicates the historical usage frequency of different exits, the zero probability $prior_{iq}$ implies that an individual *i* never uses exit q before. A possible real-life scenario is that people are told there is an exit, but they never use it previously. This case usually fits in any special passageway for fire egress or emergency use. Usually, people will not first select such an exit unless receive trustful guidance to update the probability of using it.

Now the exit choice is discrete-valued as a random number generated from the probability distribution $Prob_{iq}(t)$, which is updated by Equation (4.2). It is common to expect $Prob_{iq}(t)$ converge to a certain extent over time, but the problem exists in that the discrete-valued choices do not converge with the probability distribution. For instance, as shown in Figure 4.4, all the evacuees are given two exit choices, i.e., choosing exit A or B. Even their exit-selection distribution $Prob_{iq}(t)$ reach a consensus, for example, a distribution of [0.5, 0.5], their discrete-valued exit choices are statistically distributed evenly as half of them choose A and the other half choose B. Such a consensus in $Prob_{iq}(t)$ does not lead to a consensus in behavior. Thus, the group social force is very crucial to conquer this problem because it also critically characterizes the social relationship among individuals.

The second step is to generate the evacuation route based on the exit selected. There are a number of methods used in existing egress simulators. However, some algorithms were learned from route planning of autonomous robots, and they do not differentiate living bodies with conscious mind from non-living things like robots, and this seems a major drawback (Santos and Aguirre, 2004). In FDS+Evac the route is calculated by a two-dimensional flow solver. The computation result is a 2D flow field that guides evacuees to the exit selected. The flow field can be better explained as a social field related to social norms or other behavioral characteristics (Helbing et. al., 2005; Lewin, 1951), and we will further elaborate the idea in future. In this algorithm each exit is a sink point, and solver calculates the route as the crowd flow move to the sink (Korhonen, 2018). The detailed discussion of the flow solver will not be included in this article, but we emphasize that this method is more suitable to describe human collective behavior on the background of social science and psychological theory.

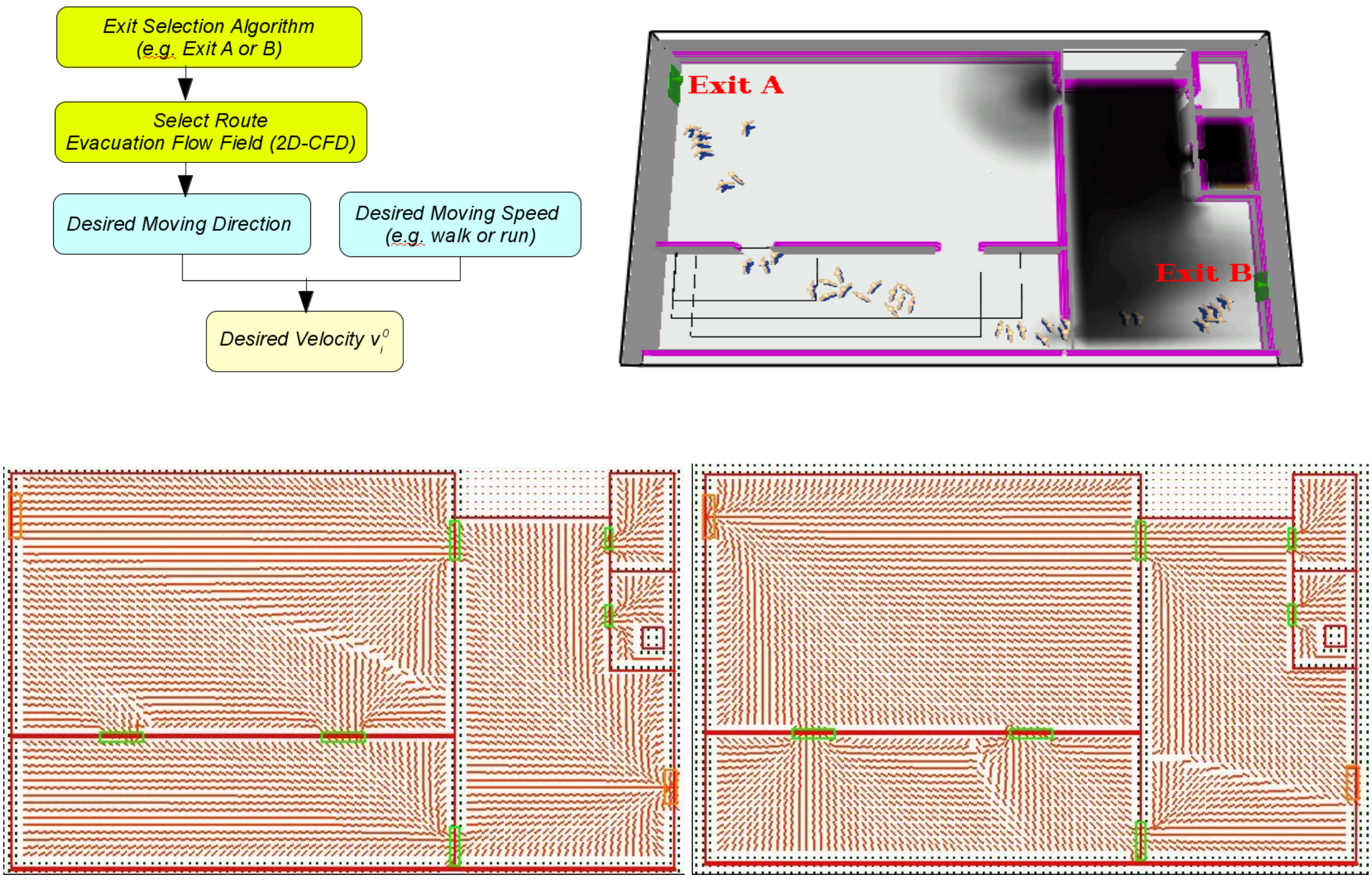

**Figure 4.5. Simulation of Crowd Evacuation with Smoke: Evacuees change their destination exit and head for the left exit. Here the desired velocity of each agent is determined by exit-selection algorithm first and their escape routes are calculated by using a 2D-computational fluid solver in FDS+Evac.**

When the target exit and route are both determined, evacuees start to move to the exit. In this process the exit and route may also be updated when new information received. For example smoke may block certain ways, and if such smoke is detected by evacuee agents, they may choose to update $\boldsymbol{v}_i^0$ to bypass smoke or select another exit or route. In case of heavy smokiness agents may directly select another exit among all the known exits and recalculate the route. A simple logic is given as below.

If Hazard_Intensity>Threshold, target exit is reselected and egress routes is recalculated.

If evacuees choose to bypass the smoke-filled area, the hazard intensity may play a major role, which is abstracted as the hazard force as mentioned in Section 2, and it determines whether people are able to move fast in smokiness. Both of the driving force and smoke resistance are increased. If the driving force is larger than the resistance, evacuees will accelerate, otherwise people have to slow down (See Figure 1.3). This algorithm has only been tested in FDS+Evac, not in CrowdEgress. Please refer to the appendix for more details.

## Conclusion

The social-force model was a physics-based model, and it was essentially developed to model either many particles or pedestrian crowd. In the past 20 years this model has been widely applied in crowd simulation in various scenarios. This article reexplains and renews the model to establish a general framework of using the model in crowd evacuation simulation. Very importantly, we introduce new concepts such as desired interpersonal distance and hazard force, and the model is thus extended in consistency of the stress theory in psychological studies, describing how environmental stressors (i.e., guidance and hazard) affect evacuee response. Such response mainly includes walking behavior at bottlenecks and smoke-filled areas. Furthermore, the pre-movement behavior and way-finding activities are discussed based on opinion dynamics and probabilistic model, and various environmental factors are integrated in this modeling framework. This simulation-based approach provides a valuable tool to observe various scenarios in crowd egress, and thus useful to identify potential risk in emergency evacuation.

Although this article specifically discusses simulating crowd behavior in evacuation conditions, we also emphasize that our modeling approach can be generalized and it contributes to many other fields. For example, the pre-evacuation dynamics and exit-choice problem can be viewed as special study cases of opinion dynamics. The force-based interactions among individuals is closely related to many-particle simulation in statistical physics and agent-based simulation in computer science and engineering. Interestingly, all the models are further linked with the stress theory in background of psychological studies and social sciences. Thus, the entire modeling framework is meaningful to bridge a major gap between engineering models, simulation practices and psychological findings in collective behavior of human crowd, and it will enable us to study a kind of complex social interaction, not only in the field of evacuation research, but also in an inter-discipline field of statistical physics and computational social science.

## Acknowledgments

This work is especially dedicated to Xiaoda Wang and Peter B. Luh for memorizing their insightful guidance in this research field. The authors thank Prof. Kerry L. Marsh, Prof. Shi-Chung Chang, Jin Sun, Bo Xiong and Vivek Kant for helpful discussion on earlier work in University of Connecticut. The authors are also thankful to Prof. Mengchu Zhou, Prof. Peter Sincak, Laura Pitukova for helpful comments on the pre-evacuation model. The author appreciates the research program funded by NSF Grant # CMMI-1000495 (NSF Program Name: Building Emergency Evacuation - Innovative Modeling and Optimization).

## Supplementary data

The supplementary data to this article are available online at https://sourceforge.net/p/crowdegress or https://github.com/godisreal/crowdEgress. The output data of FDS+Evac examples are uploaded in the Github repository https://github.com/crowdmodel/test-crowd-dynamics. For instance there are standard example of a single room with two exits or three exits, which are basically used to test how agents socially interact and select different exits (See Figure 4.4 and Figure 4.5). The example of a room with one exit is also widely used to test crowd behavior at bottlenecks such as the faster-is-slower effect (Helbing, Farkas, Vicsek, 2000). There are also some more complicated examples. A typical one is extended from MassEgress project (Pan, 2006), as shown in Figure 3.3 and Figure 3.4, and another one is from our IEEE Conference paper (Wang et. al., 2008) as shown in Figure 3.5.

Other numerical testing cases are included online at https://sourceforge.net/p/socialnetwork-abs and https://github.com/crowdmodel/group-social-force. Provided that you have any comment or inquiry about the testing results, please feel free to send emails to wp2204@gmail.com or start an issue on the repository.

## APPENDIX

### (a) Hazardous Effect on Pedestrian Movement

In building egress whether people can move fast critically depends on hazard intensity. As illustrated in Figure 1.3 we provide an example for quantitative analysis of hazard effect based on Equation (1.1). For example, in thin smoke people are able to speed up while heavy smokiness impede their motion significantly. An example is that $\boldsymbol{f}_{ih}$ increases in a quadratic rare of smoke density while the driving force increases in a square root form or a linear form of smoke density (See Figure 1.3). The above approach has been partly implemented and tested in FDS+Evac as below.

In the original setting of FDS+Evac an evacuation process is stimulated by using a pedestrian model extended from social-force model, where the psychological desire of motion is described by desired velocity $v^0$. The desired velocity $v^0$ is next coupled with the fire/smoke condition: In a non-smoke area $v^0$ is equal to a preset value called the unimpeded walking speed, which is human walking speed in normal situation (e.g., 0.8m/s - 1.6m/s). If smoke density increases, $v^0$ will increase correspondingly, e.g., $v_{\mathrm{i}}^0(\mathrm{t}) = v_{\mathrm{i}}^{\text{0-preset}}(\mathrm{t}) + K_1$ (SOOT_DENS 1)$^{1/2}$. As a result, larger $v^0$ is assigned if the agent perceives or detects the smoke in the environment, and thus desire moving faster for escape. This setting leads to an increase of driving force. Especially, we measure such increase by the force component $(m_i\boldsymbol{v}_i^0)/\tau$ in the driving force of $\boldsymbol{f}^{drv} = m_i(\boldsymbol{v}_i^0 - \boldsymbol{v}_i)/\tau$.

In contrast the hazard force is specified by $|\boldsymbol{f}_{ih}|=K_2$*SOOT_DENS $^2$ to impede an evacuee's physical motion even if the agents is exactly moving in a smoke-filled area. Here the formula of hazard force is selected mainly due to Stevens's power function on human perception of stimuli (Stevens, 1971), where *SOOT_DENS* represents the physics measure of smoke density, and human perception of smoke is described by $K_2$*SOOT_DENS$^2$. In a general sense $|\boldsymbol{f}_{ih}|$ is denoted by high-order power function of SOOT_DENS or other hazard intensity, and this formula describes that people become sensitive to hazard intensity if it exceeds a certain threshold, and this mathematical description is generally consistent with existing psychological findings.

HR%FX_Hazard and HR%FY_Hazard are the force elements added to HUMAN_TYPE. SMOKE_BLK_FAC is a damping coefficient which slows down agents' movement when agents walk in smoke condition. HEAT_GRAD_FAC is a parameter which tunes smoke resistance with respect to gradient of gas temperature TMP_G.

```
HR%FX_Hazard = –HEAT_GRAD_FAC*(HUMAN_GRID(II,JJ)%TMP_G – HUMAN_GRID(II-1,JJ)%TMP_G)*HUMAN_GRID(II,JJ)%TMP_G**2 – SMOKE_BLK_FAC*HR%U*HUMAN_GRID(II,JJ)%SOOT_DENS**2/SQRT(HR%U**2 + HR%V**2)

HR%FY_Hazard = –HEAT_GRAD_FAC*(HUMAN_GRID(II,JJ)%TMP_G – HUMAN_GRID(II,JJ-1)%TMP_G)*HUMAN_GRID(II,JJ)%TMP_G**2 – SMOKE_BLK_FAC*HR%V*HUMAN_GRID(II,JJ)%SOOT_DENS**2/SQRT(HR%U**2 + HR%V**2)

HR%FX_Hazard = min(HR%FX_Hazard, HR%Mass*2.0_EB)    ! Give a lower bound of hazard force

HR%FY_Hazard = min(HR%FY_Hazard, HR%Mass*2.0_EB)    ! Give a lower bound of hazard force
```

### (b) Group Social Force by Fortran Code in FDS+Evac

```
!! Group social force is added. Here I declare a 2D matrix when the number of EVAC lines is determined. Users may initialize or modify the matrix by using EVAC Namelist, and this matrix characterizes the social relationship of agents.
!! The total number of groups is NPC_EVAC, then DFAC AFAC and BFAC are in dimension of NPC_EVAC*NPC_EVAC.
!! DFAC: DFactor(I, IE)    AFAC: AFactor(I, IE)       BFAC: BFactor(I, IE)
!! I: Index of the current agent (outer loop)   IE: Index of the other agent (inner loop)
!! The following Fortran code is in the loop where social force is computed.
!!  GROUP_FORCE is a boolean variable which enables the group dynamics in computation.

          FCG_X = 0.0_EB
```

```
          FCG_Y = 0.0_EB
          IF (GROUP_FORCE) THEN
             TIM_DIST = MAX(0.001_EB,SQRT((X_TMP(2)-X_TMP(5))**2 + (Y_TMP(2)-Y_TMP(5))**2))
                    FCG_X    =    (X_TMP(2)-X_TMP(5))*HR_A*AFAC*COSPHIFAC*EXP(    -(TIM_DIST-
( R_TMP(2)+R_TMP(5) )
     *DFAC)/HR_B/BFAC)/TIM_DIST*(( R_TMP(2)+R_TMP(5) )*DFAC-TIM_DIST)
                    FCG_Y    =    (Y_TMP(2)-Y_TMP(5))*HR_A*AFAC*COSPHIFAC*EXP(    -(TIM_DIST-
( R_TMP(2)+R_TMP(5) )
     *DFAC)/HR_B/BFAC)/TIM_DIST*(( R_TMP(2)+R_TMP(5) )*DFAC-TIM_DIST)
          END IF

          !HR_A_CF
          !HR_B_CF
          HR%FX_Group = HR%FX_Group + FCG_X
          HR%FY_Group = HR%FY_Group + FCG_Y
```

### (c) Analytic Hierarchy Process (AHP) to determine the parameter a1, a2, a3

Basically we investigate how people weigh different factors in their decision making processes. Suppose a person considers other people's decision weighs 3 times of importance as their prior knowledge, and their prior knowledge is 2 times important as their current investigation result. The following AHP matrix is given to describe the case. As commonly known, improper user input at <?> positions in AHP matrix could cause inconsistency problem.

| $AHP$ | $a1$ | $a2$ | $a3$ |
|---|---|---|---|
| $a1$ | 1 | 1/2 | 3 |
| $a2$ | 2 | 1 | ? |
| $a3$ | 3 | ? | 1 |

Here we will use a kind of hierarchical approach rather than the standard AHP approach to avoid inconsistency problem. That is we will firstly determine $a_{12}=(a_1+a_2)$ and $a_3$, where $(a_1+a_2)$ represents the tendency of people to make a decision by themselves and $w_3$ denotes the tendency of following others decision. Based on AHP techniques the weigh matrix is computed in two steps as introduced below.

Step1: Suppose one think others' decision is three times of importance as his or her individual decision. The AHP array is specified as below.

| $AHP$ | $a12$ | $a3$ |
|---|---|---|
| $a12$ | 1 | 3 |
| $a3$ | 1/3 | 1 |

$$\begin{bmatrix} 1 & 3 \\ 1/3 & 1 \end{bmatrix} \text{ --Normalize}\rightarrow \begin{bmatrix} 0.75 & 0.75 \\ 0.25 & 0.25 \end{bmatrix}$$

Step2: Suppose one consider that his or her prior experience is two times of importance as investigating current situation. The AHP array is thus given by

| $AHP$ | $a1$ | $a2$ |
|---|---|---|
| $a1$ | 1 | 1/2 |
| $a2$ | 2 | 1 |

$$\begin{bmatrix} 1 & 1/2 \\ 2 & 1 \end{bmatrix} \text{ --Normalize}\rightarrow \begin{bmatrix} 1/3 & 1/3 \\ 2/3 & 2/3 \end{bmatrix}$$